\documentclass[twocolumn,showpacs,preprintnumbers,superscriptaddress,amsmath,amssymb,aps,pre]{revtex4-1}
\usepackage{epsfig}
\usepackage{graphicx}
\usepackage{dcolumn}
\usepackage{bm}
\usepackage{times}
\usepackage{xcolor}
\usepackage{booktabs}
\usepackage{subfigure} 

\begin{document}

\title{Optimal networks for dynamical spreading}

\author{Liming Pan}
\affiliation{School of Computer and Electronic Information, Nanjing Normal University, Nanjing, Jiangsu, 210023, China}

\author{Wei Wang} \email{wwzqbx@hotmail.com}
\affiliation{Cybersecurity Research Institute, Sichuan University, Chengdu 610065, China}

\author{Lixin Tian}
\affiliation{School of Mathematical Sciences, Nanjing Normal University, Nanjing, Jiangsu 210023, China}

\author{Ying-Cheng Lai} 
\affiliation{School of Electrical, Computer and Energy Engineering, Arizona State University, Tempe, Arizona 85287, USA}
\affiliation{Department of Physics, Arizona State University, Tempe, Arizona 85287, USA}

\date{\today}

\begin{abstract}

The inverse problem of finding the optimal network structure for a specific type of dynamical process stands out as one of the most challenging problems in network science. Focusing on the susceptible-infected-susceptible type of dynamics on annealed networks whose structures are fully characterized by the degree distribution, we develop an analytic framework to solve the inverse problem. We find that, for relatively low or high infection rates, the optimal degree distribution is unique, which consists of no more than two distinct nodal degrees. For intermediate infection rates, the optimal degree distribution is multitudinous and can have a broader support. We also find that, in general, the heterogeneity of the optimal networks decreases with the infection rate. A surprising phenomenon is the existence of a specific value of the infection rate for which any degree distribution would be optimal in generating maximum spreading prevalence. The analytic framework and the findings provide insights into the interplay between network structure and dynamical processes with practical implications.

\end{abstract}
\maketitle

\section{Introduction} \label{sec:intro}

In the study of dynamics on complex networks, most previous efforts were 
focused on the {\em forward} problem: How does the network structure affect 
the dynamical processes on the network? The approaches undertaken to address
this question have been standard and relatively straightforward: One implements
the dynamical process of interest on a given network structure and then 
studies how alterations in the network structure affect the dynamics. The 
{\em dynamical inverse} problem is much harder: finding a global network 
structure that optimizes a given type of dynamical processes. Despite the extensive and intensive efforts in the past that have resulted in 
an essential understanding of the interplay between dynamical processes and 
network structure, previous studies of the inverse problem were sporadic 
and limited to a perturbation type of analysis, generating solutions that are 
at most locally optimal only~\cite{aguirre2013successful,pan2019optimal}. 
The purpose of this paper is to present and demonstrate an analytic framework 
to address the dynamical inverse problem. 

To be concrete, we will focus on spreading dynamics on networks for which
a large body of literature has been generated in the past on the forward 
problem, i.e., how network topology affects the characteristics of the 
spreading, such as the outbreak threshold and 
prevalence~\cite{pastor2015epidemic,castellano2009statistical}. For example, 
under the annealed assumption that all nodes with the same degree are 
statistically equivalent, it was found~\cite{pastor2001epidemic} that the 
epidemic threshold of the susceptible-infected-susceptible (SIS) process is 
given by $\langle k\rangle/\langle k^2\rangle$, where $\langle k\rangle$ and 
$\langle k^2\rangle$ are the first and second moments of the degree 
distribution, respectively. In situations where the second moment diverges,
the threshold value is essentially zero, meaning that the presence of a few
hub nodes can greatly facilitate the occurrence of an epidemic outbreak.
An understanding of the interplay between the network structure and the spreading
dynamics is essential to articulating control strategies. For example, 
the important role played by the hub nodes suggests a mitigation strategy:
Vaccinating these nodes can block or even stop the spread of the 
disease~\cite{cohen2003efficient,pastor2002immunization}. Likewise, if the 
goal is to promote information spreading, then choosing the hub nodes as  
the initial seeds can be effective~\cite{kitsak2010identification,lu2016vital}.

The inverse problem is motivated by the application scenarios in which one
strives to optimize the network structure to achieve desired or improved
performance~\cite{valente2012network}. Optimization and invention have been
applied to problems such as virus marketing~\cite{goel2016structural}, social 
robots detection~\cite{ferrara2016rise}, containment of false news 
spreading~\cite{vosoughi2018spread}, and polarization reduction in social 
networks~\cite{musco2018minimizing}. For spreading dynamics on networks, 
the few existing studies are focused on applying small perturbations to the 
network structure to modulate the dynamical 
process~\cite{aguirre2013successful,pan2019optimal}. From the point of view 
of optimization, since the perturbations are local, the resulting solution is 
locally optimal at best. 

We address the following questions: Does a globally optimal network exist
and if yes, can it be found to maximize the prevalence of the spreading 
dynamics? Such a network is necessarily extremum. 
For general types of spreading dynamics, to analytically solve this inverse
problem is currently not feasible. However, we find that the SIS type of 
spreading dynamics does permit an analytic solution. In particular, the annealed
approximation stipulates that the network structure can be fully captured or 
characterized by its degree distribution. The problem of finding the optimal 
networks can then be formulated as one to find the optimal degree distribution 
that maximizes the prevalence of the SIS spreading dynamics, which can be
analytically solved by exploiting the heterogeneous mean-field (HMF) 
theory~\cite{pastor2015epidemic}. Notwithstanding the necessity of imposing 
the annealed approximation to enable analytic solutions, the essential 
physical ingredients of the SIS dynamics are retained. 

Our main results are the following. Taking a variational approach to solving 
the HMF equation, we obtain a necessary condition for the optimal degree 
distribution. The condition defines a set of candidate optimal degree 
distributions, and we show that a degree distribution is globally optimal if 
and only if it belongs to the set. However, if the set is empty, which can 
occur for relatively low and high infection rates, the necessary condition 
stipulates that a local extremum distribution must concentrate on no more 
than two distinct nodal degree values thereby substantially narrowing the 
search for the optimal network. Searching through all possible distributions 
under the constraint leads to the optimal degree distribution that can be 
proved to be unique. For intermediate infection rates, multiple optimal degree 
distributions with a broader support exist, which lead to identical spreading 
prevalence. In addition, our theory predicts the existence of a particular 
value of the infection rate for which every degree distribution is optimal. 
A general trend is that the degree heterogeneity of the optimal distribution 
decreases with the infection rate. 

Our paper represents a first step toward finding a global optimal network 
structure for spreading dynamics. From a theoretical point of view, developing 
a method to find such extremum networks represents a feat that would provide 
deeper insights into the interplay between network topology and spreading 
dynamics. From a practical perspective, the solution can be exploited to 
design networks that are capable of spreading information or transporting 
material substances in the most efficient way possible. 

In Sec.~\ref{sec:model}, we introduce the HMF theory for the SIS dynamics 
and set up the basic framework for the optimization problem. In 
Sec.~\ref{sec:necess_cond}, we employ a variational method to derive the 
necessary condition for a degree distribution to be an extremum among all 
feasible distributions. Solutions of the optimal degree distribution are 
presented in Sec.~\ref{sec:find_optim}, and its properties are discussed in 
Sec.~\ref{sec:properties}. The paper is concluded in Sec.~\ref{sec:discussion} 
with a discussion. 

\section{Problem formulation and sketch of major mathematical steps} \label{sec:model}

In the SIS model, each node can be either in the susceptible or in the 
infected state, and we assume the nodal state evolves continuously with time. 
During the spreading process, a susceptible node is infected by its neighbors 
with the rate $\lambda$, whereas an infected node recovers at the rate 
$\gamma$. To study the equilibrium properties of the dynamical process, it 
is convenient to set $\gamma=1$ so that $\lambda$ is the sole dynamical parameter.

In the HMF theory, all the nodes with the same degree are statistically 
equivalent~\cite{pastor2015epidemic}. Consider a vector of nodal degrees 
$\mathbf{k}\equiv [k_1,k_2,\cdots,k_n]^T$, where the elements are arranged 
in a descending order: $k_1>k_2>\cdots>k_n$. The degree distribution is fully
specified by a probability vector defined as 
$\mathbf{p}\equiv [p_1,p_2,\cdots,p_n]^T$, where 
$p_i\geq 0$ is the probability that a randomly chosen node has degree $k_i$. 
Let $x_i(t)$ be the probability that a node with degree $k_i$ is infected at 
time $t$. Given the probability vector $\mathbf{p}$, the HMF equation is 
\begin{equation} \label{eq:HMF}
\frac{d x_i(t)}{dt}= - x_i(t)+\lambda k_i \left[1-x_i(t)\right]\Theta
\end{equation}
for $i\in\{1,\cdots,n\}$, where
\begin{equation}\label{eq:Theta}
\Theta=\frac{1}{\langle k\rangle} \sum_{j=1}^n p_j k_j x_{j} (t).
\end{equation}

In Ref.~\cite{wang2008global}, it was proved that the HMF equation has a 
unique global stable equilibrium point $x^*$.  In addition, for 
$\lambda < \langle k \rangle/\langle k^2\rangle$, we have $x^*_i=0$ for all 
$i\in\{1,\cdots,n\}$, whereas for $\lambda>\langle k \rangle/\langle k^2\rangle$,
we have $0<x^*_i<1$ for all $i\in \{1,\cdots,n\}$. The spreading prevalence 
in the equilibrium state is
\begin{equation}
	\psi(\mathbf{p})=\sum_{i=1}^{n} p_i x_i^*,
\end{equation}
where, to simplify the notations, we have omitted the dependence of 
$\psi(\mathbf{p})$ on $\lambda$. Let $\mathcal{P}$ be the family of all 
degree distributions with a fixed average degree defined on $\mathbf{k}$. 
That is, with a prespecified constant $z>0$, for any 
$\mathbf{p} \in\mathcal{P}$, we have $\sum_{i=1}^n p_i k_i=z$. Our goal is 
to find $\mathbf{p}^o\in \mathcal{P}$ that maximizes $\psi(\mathbf{p})$: 
\begin{equation}\label{eq:opt_pbl}
\mathbf{p}^o=\underset{\mathbf{p}\in \mathcal{P}}{\mathrm{argmin}}\ \psi(\mathbf{p}).
\end{equation}
The optimization problem is nontrivial only when the value of $\lambda$ is 
larger than the epidemic threshold at least for one $\mathbf{p}\in\mathcal{P}$.
The Bhatia-Davis inequality stipulates that the second moment of $\mathbf{p}$ 
is maximized when $\mathbf{p}$ concentrates on the end points $k_1$ and $k_n$. 
In this case, the second moment is $\langle k^2\rangle=zk_1+zk_n-k_1k_n$. The 
optimization problem is nontrivial only when the following condition is met:
\begin{equation} \label{eq:lambda1_def}
\lambda>\lambda_1 \equiv \frac{z}{zk_1+zk_n-k_1k_n}.
\end{equation} 
In this case, if there is a unique solution $\mathbf{p}$ such that 
$\lambda >z/\langle k^2\rangle$, it gives the optimal degree distribution 
$\mathbf{p}^o$.

Our goal is to analytically find the solutions for the optimization
problem defined in Eq.~\eqref{eq:opt_pbl}. As the mathematical derivations 
involved are lengthy, it may be useful to sketch the basic idea, tools used,
and the results, which we organize as the following three major steps.
\begin{enumerate}
\item
Mathematically, Eq.~\eqref{eq:opt_pbl} defines a variational problem for the 
HMF equations in Eq.~\eqref{eq:HMF}, which can be studied through the standard
calculus-of-variation techniques. In Sec.~\ref{subsec:variation}, we adopt 
a variational approach for the HMF equations in Eq.~(1) and derive the 
necessary condition for a degree distribution to be optimal. In particular, 
we impose a perturbation to the degree distribution as
$\mathbf{p}^{\prime}=\mathbf{p}+\alpha \bar{\mathbf{p}}$ and derive a formula
that predicts $\bar{\psi}(\mathbf{p},\mathbf{p}^{\prime})$, the part of the
incremental spreading prevalence which is linear in $\alpha$. For $\mathbf{p}$
to be a candidate maximum, $\bar{\psi}(\mathbf{p},\mathbf{p}^{\prime})$ must be
nonpositive for any choice of $\bar{\mathbf{p}}$, and this leads to the necessary condition for the local minima.
\item
The next task is to study the necessary condition resulting from the 
variational analysis. In Sec.~\ref{subsec:consequences}, through a sequence 
of algebraic arguments, we show that for any $\mathbf{p}$ satisfying the 
necessary condition, it is only possible to have either
(\romannumeral1) $\bar{\psi}(\mathbf{p},\mathbf{p}^{\prime})= 0$ or
(\romannumeral2) $\bar{\psi}(\mathbf{p},\mathbf{p}^{\prime})<0$ for all
feasible perturbations. This means that it is impossible to find a certain
$\mathbf{p}$ such that $\bar{\psi}(\mathbf{p},\mathbf{p}^{\prime})= 0$ and
$\bar{\psi}(\mathbf{p},\mathbf{p}^{\prime\prime})<0$ for
$\mathbf{p}^{\prime}\neq \mathbf{p}^{\prime\prime}$. Further, in
Sec.~\ref{subsec:consequences}, we show that the condition
$\bar{\psi}(\mathbf{p},\mathbf{p}^{\prime})= 0$ can be reduced to a linear
equation in $\mathbf{p}$ [the first equation in \eqref{eq:po_def}] which,  
together with the probability constraint $\sum_{i=1}^n p_i=1$ 
and the average degree constraint $\sum_{i=1}^n p_i k_i=z$, defines a set of 
candidate optimal degree distributions $\mathcal{P}^o$. In 
Sec.~\ref{subsec:global}, by analyzing the three linear equations, we show 
that if $\mathcal{P}^o$ is nonempty, then any $\mathbf{p}$ is a global maximum 
if and only if $\mathbf{p}\in \mathcal{P}^o$. Concurrently, if $\mathcal{P}^o$ 
is empty, the optimal degree distribution with 
$\bar{\psi}(\mathbf{p},\mathbf{p}^{\prime})<0$ will concentrate on no more 
than two distinct nodal degrees.
\item
Finally, in Sec.~\ref{subsec:nonempty}, we derive the condition when the set
$\mathcal{P}^o$ is nonempty by analyzing the three linear equations defining
the set [Eq.~(25)]. In particular, $\mathcal{P}^o$ is nonempty for
$\lambda\in [\lambda_2,\ \lambda_3]$ (see Sec.~\ref{subsec:nonempty} for 
explicit definitions of $\lambda_2$ and $\lambda_3$). For $\lambda<\lambda_2$ 
or $\lambda>\lambda_3$ and $\mathcal{P}^o$ indeed empty, we find the optimal
degree distributions by solving the HMF equations explicitly 
(Sec.~\ref{subsec:two_solution}).
\end{enumerate}

\section{Necessary condition for local extrema and consequences} \label{sec:necess_cond}

In this section, we first study the optimization problem defined in 
Eq.~\eqref{eq:opt_pbl} using several techniques from the calculus of variation. The calculation provides a necessary condition for finding the local maxima. 
We then analyze the necessary condition in detail to find the global optimal 
degree distributions.

\subsection{Variational method} \label{subsec:variation}

We study the variation problem in Eq.~\eqref{eq:opt_pbl} using the standard 
techniques from the calculus of variations. Briefly, we apply a perturbation 
to the degree distribution $\mathbf{p}$ in Eq.~\eqref{eq:HMF} and calculate the 
linear response for the spreading prevalence. A local maximum necessarily has 
non-positive linear responses for any feasible perturbation.

For a fixed $\lambda>\langle k \rangle/\langle k^2\rangle$, let $x^*$ be the corresponding 
globally stable equilibrium point of the HMF equation. We impose a small 
variation on $p_i$,
\begin{equation}
p^{\prime}_i = p_i + \alpha\bar{p}_i,
\end{equation}
where $\bar{\mathbf{p}}$ specifies the direction of the variation and 
$\alpha>0$ controls its magnitude. For the perturbed degree distribution to 
be feasible, i.e., $\mathbf{p}^{\prime}\in\mathcal{P}$, the following 
conditions are necessary:
\begin{equation} \label{eq:p_constraint}
	\sum_{i=1}^n \bar{p}_i=0 \ \ \hbox{and} \ \ \sum_{i=1}^n \bar{p}_i k_i=0.
\end{equation}
In addition, the perturbed degree distribution $\mathbf{p}^{\prime}$ must
satisfy the probability constraints $0\leq p^{\prime}_i\leq 1$.

Let $x^{\prime}(t,\alpha)$ be the trajectory of the perturbed system. 
The time evolution of $x^{\prime}(t,\alpha)$ is described by the HMF equation with $\mathbf{p}$ replaced by $\mathbf{p}^{\prime}$ and $x_i(t)$ in 
Eq.~\eqref{eq:HMF} by $x_i^{\prime}(t,\alpha)$. As shown in 
Appendix~\ref{appendix_A}, $\mathbf{x}^*$ is a continuously differentiable 
function of $\mathbf{p}$ for $\lambda > z/\langle k^2\rangle$, enabling 
the following expansion of $x^{\prime}(t,\alpha)$ about $x^*_i$: 
\begin{equation} \label{eq:x_expand}
x^{\prime}_i(t,\alpha)=x^*_i+\alpha \bar{x}_i(t)+o(\alpha),
\end{equation}
where $\bar{x}_i(t)$ is the response to the perturbation which is linear in 
$\alpha$. Taking the derivative with respect to $\alpha$ at $\alpha=0$, we 
obtain  
$\partial x^{\prime}_i(t,\alpha)/\partial \alpha \vert_{\alpha=0}=\bar{x}_i(t)$.
The time derivative of $\bar{x}_i(t)$ is then given by
\begin{equation}
\begin{split}
\frac{d \bar{x}_i(t)}{dt}&=\frac{\partial}{\partial \alpha}\bigg|_{\alpha=0} \frac{d x_i(t,\alpha)}{dt},
\end{split}
\end{equation} 
which, after some algebraic manipulations, can be rewritten as
\begin{equation}\label{eq:bx_t}
\frac{d \bar{\mathbf{x}}(t)}{dt}=\mathcal{J} \bar{\mathbf{x}}(t)+\xi,
\end{equation}
where $\mathcal{J}$ is the $n\times n$ Jacobian matrix that does not depend on 
$\bar{\mathbf{p}}$ and $\xi$ is a vector of length $n$ that depends on 
$\bar{\mathbf{p}}$. The elements of $\mathcal{J}$ and $\xi$ are given by
\begin{equation} \label{eq:X}
J_{ij}=-\delta_{i,j}\left(1+\lambda k_i \Theta ^*\right)+\frac{\lambda}{z} k_i\left(1-x^*_i\right)k_j p_j,
\end{equation}
and
\begin{equation}
\xi_i=\frac{\lambda}{z} k_i \left(1-x^*_i\right)\sum_{j=1}^n k_j \bar{p}_j x^*_j,
\end{equation}
respectively. In Eq.~\eqref{eq:X}, $\delta_{i,j}$ is the Kronecker $\delta$ and 
$\Theta^*$ is obtained by substituting $\mathbf{x}(t)=\mathbf{x}^*$ into 
Eq.~\eqref{eq:Theta}.

Equation~\eqref{eq:bx_t} defines a linear system with the solution,
\begin{equation}
\bar{\mathbf{x}}(t)=\mathrm{e}^{\mathcal{J}t} \bar{\mathbf{x}}(0)+\left(\mathrm{e}^{\mathcal{J}t}-\mathcal{I}\right) \mathcal{J}^{-1} \xi,
\end{equation}
where $\mathrm{e}^{\mathcal{J}t}$ is the matrix exponential of $\mathcal{J}t$. 
In Appendix~\ref{appendix_B}, we show that the eigenvalues of $\mathcal{J}$ 
have negative real parts. In the long time limit, we then have
\begin{equation} \label{eq:bx_star}
\bar{\mathbf{x}}^*=\lim_{t\to \infty}\bar{\mathbf{x}}(t)=-\mathcal{J}^{-1}\xi.
\end{equation}
With the perturbed degree distribution and Eq.~\eqref{eq:x_expand}, we can
express the spreading prevalence as
\begin{equation}
\psi(\mathbf{p}^{\prime})=\psi(\mathbf{p})+\alpha \bar{\psi}\left(\mathbf{p},\mathbf{p}^{\prime}\right)+o(\alpha),
\end{equation}
where $\bar{\psi}(\mathbf{p},\mathbf{p}^{\prime})$ is the part of incremental 
spreading prevalence that is linear in $\alpha$,
\begin{equation}\label{eq:bpsi}
\bar{\psi}(\mathbf{p},\mathbf{p}^{\prime})=\sum_{i=1}^n\left(\bar{p}_i x^*_i+p_i\bar{x}^*_i\right).
\end{equation}
Substituting Eq.~\eqref{eq:bx_star} into Eq.~\eqref{eq:bpsi}, we have (after
some algebraic manipulations) 
\begin{equation} \label{eq:bpsi_2}
\bar{\psi}(\mathbf{p},\mathbf{p}^{\prime})=\sum_{i=1}^n\chi_i \bar{p}_i,
\end{equation}
where $\chi_i$ is given by
\begin{equation}\label{eq:chi}
\chi_i=x^*_i\left(1+\frac{\lambda k_i \Theta^* \sum_{j=1}^n p_j k_j(1-x^*_j)^2}{\sum_{j=1}^n p_j k_j(x^*_j)^2}\right),
\end{equation}
The detailed derivation of Eq.~\eqref{eq:chi} is presented in 
Appendix~\ref{appendix_C}. The \emph{necessary condition} for the degree distribution
$\mathbf{p}$ to be a local maximum is if and only if the inequality 
$\bar{\psi}(\mathbf{p},\mathbf{p}^{\prime})\leq 0$ holds for all feasible 
perturbations.

\subsection{Consequences of the necessary condition} \label{subsec:consequences}

Equations~\eqref{eq:bpsi_2} and \eqref{eq:chi} allow us to significantly 
narrow the search range for the optimal degree distribution through the 
process of elimination. In the following, we analyze the necessary condition 
by proving that it is only possible to have either
(\romannumeral1) $\bar{\psi}(\mathbf{p},\mathbf{p}^{\prime})= 0$ or
(\romannumeral2) $\bar{\psi}(\mathbf{p},\mathbf{p}^{\prime})<0$ for all
feasible perturbations. That is, it is impossible to find 
$\mathbf{p}$ such that $\bar{\psi}(\mathbf{p},\mathbf{p}^{\prime})= 0$ and
$\bar{\psi}(\mathbf{p},\mathbf{p}^{\prime\prime})<0$ for
$\mathbf{p}^{\prime}\neq \mathbf{p}^{\prime\prime}$. We then show that
$\bar{\psi}(\mathbf{p},\mathbf{p}^{\prime})= 0$ can be reduced to an equation
that is linear in $\mathbf{p}$, based on which the spreading prevalence for
any $\mathbf{p}$ satisfying $\bar{\psi}(\mathbf{p},\mathbf{p}^{\prime})= 0$
can be directly obtained without solving the HMF equations. The results in
this section are obtained through algebraic manipulations of the equation
$\bar{\psi}(\mathbf{p},\mathbf{p}^{\prime})= 0$.

The starting point of our analysis is to determine when the linear variation 
$\bar{\psi}(\mathbf{p},\mathbf{p}^{\prime})$ vanishes. A feasible perturbation 
$\bar{\mathbf{p}}$ must satisfy the constraints in \eqref{eq:p_constraint},
so $\bar{\mathbf{p}}$ must have at least three nonzero elements. Pick any 
$m\geq 3$ points $\{k_{i_1},k_{i_2},\cdots k_{i_m}\}$ from $\mathbf{k}$ and 
consider a perturbation $\bar{\mathbf{p}}$ whose elements are nonzero only on 
these points. The linear variation $\bar{\psi}(\mathbf{p},\mathbf{p}^{\prime})$
vanishes only if $\mathcal{Z}^{(m)}\bar{\mathbf{p}}=0$, where 
$\mathcal{Z}^{(m)}$ is a $3\times m$ matrix,
\begin{equation}
\mathcal{Z}^{(m)}=\left(
\begin{array}{cccc}
1 & 1 & \cdots & 1\\
k_{i_1} & k_{i_2} & \cdots & k_{i_m}\\
\chi_{i_1} & \chi_{i_2} & \cdots & \chi_{i_m}
\end{array}\right).
\end{equation} 
The first two rows in $\mathcal{Z}^{(m)}$ correspond to the constraints for 
$\bar{\mathbf{p}}$ in \eqref{eq:p_constraint}, while the last row 
is the result of the definition of $\bar{\psi}(\mathbf{p},\mathbf{p}^{\prime})$
in Eq.~\eqref{eq:bpsi_2}. To gain insights, we temporally disregard the 
probability constraint $\mathbf{p}^{\prime}\in [0\ 1]^n$ (which will be 
included in the analysis later). Under this condition, any $\bar{\mathbf{p}}$ 
that makes the linear variation $\bar{\psi}(\mathbf{p},\mathbf{p}^{\prime})$ 
vanish belongs to the null space of $\mathcal{Z}^{(m)}$. By the rank-nullity 
theorem, we have 
$\mathrm{nullity}(\mathcal{Z}^{(m)})=m-\mathrm{rank}(\mathcal{Z}^{(m)})$. 
The dimension of the space for all feasible perturbations, i.e., the nullity 
of the sub-matrix consisting of the first two rows of $\mathcal{Z}^{(m)}$, is 
$m-2$. As a result, the linear variation vanishes for all directions of 
perturbation if $\mathrm{nullity}(\mathcal{Z}^{(m)})=m-2$, which further 
implies the condition $\mathrm{rank}(\mathcal{Z}^{(m)})=2$. We thus have that
the linear variation vanishes if and only if the third row of 
$\mathcal{Z}^{(m)}$ is a linear combination of the first two rows.

Setting the right-hand-side of Eq.~\eqref{eq:HMF} to zero, we obtain the 
equilibrium solution as $x^*_i=\lambda k_i \Theta^*/(1+\lambda k_i \Theta^*)$. 
From the definition of $\chi_i$ in Eq.~\eqref{eq:chi}, we have
\begin{equation}
\chi_i=\frac{\lambda k_i}{1+\lambda k_i \Theta^*}\left(1+\frac{\lambda \Theta^* k_i \sum_{j=1}^n p_j k_j(1-x^*_j)^2}{\sum_{j=1}^n p_j k_j(x^*_j)^2}\right).
\end{equation}
If the following holds
\begin{equation}\label{eq:local_label}
\frac{\sum_{j=1}^n p_j k_j(1-x^*_j)^2}{\sum_{j=1}^n p_j k_j(x^*_j)^2}=1,
\end{equation}
then we have $\chi_i=\lambda k_i$. In this case, the third row of 
$\mathcal{Z}^{(m)}$ is exactly the second row multiplying by $\lambda$ and we 
have $\mathrm{rank}(\mathcal{Z}^{(m)})=2$. Moreover, if 
Eq.~\eqref{eq:local_label} holds, then $\mathrm{rank}(\mathcal{Z}^{(m)})=2$ 
holds for any choice of perturbation with $m\geq 3$. In other words, the linear variation thus vanishes for all 
directions of perturbation.

In the above analysis, we have not required 
$\mathbf{p}+\alpha \bar{\mathbf{p}}\in [0\ 1]^n$. A direction of perturbation 
$\bar{\mathbf{p}}$ would be infeasible if an element of $\mathbf{p}$ has 
$p_i=0$ or $p_i=1$. Nevertheless, as Eq.~\eqref{eq:local_label} guarantees 
$\mathcal{Z}^{(m)} \bar{\mathbf{p}}=0$ for any $m$, the linear variation 
$\bar{\psi}(\mathbf{p},\mathbf{p}^{\prime})$ vanishes in any direction of 
perturbation, feasible or infeasible. In fact, in the proof of 
$\mathbf{x}^*$ being a continuously differentiable function of $\mathbf{p}$ 
(Appendix~\ref{appendix_A}), it is not necessary to require $p_i\neq 0$ or 
$p_i\neq 1$ for any $i\in \{1,\cdots,n\}$. This means that the perturbation 
in an infeasible direction can still be well-defined, although it is physically
irrelevant. Consequently, Eq.~\eqref{eq:local_label} provides the sufficient 
condition for a local extremum.

The analysis so far gives that a local maximum of $\psi(\mathbf{p})$ either 
has: (\romannumeral1) $\bar{\psi}(\mathbf{p},\mathbf{p}^{\prime})= 0$ in any 
direction of perturbation, or (\romannumeral2) 
$\bar{\psi}(\mathbf{p},\mathbf{p}^{\prime})<0$ for all feasible perturbations. 
It is not possible to find a local maximum such that the linear variation 
vanishes in some directions of perturbation and negative in others. Notice 
that case (\romannumeral1) only provides a necessary condition for a local 
extremum and we need to further determine if it is a maximum or a minimum.

To proceed, we continue to analyze the local extrema with 
$\bar{\psi}(\mathbf{p},\mathbf{p}^{\prime})= 0$ from 
Eq.~\eqref{eq:local_label} which, for $x^*_i> 0$, can be rewritten as
\begin{equation}
\sum_{j=1}^n p_j k_j=2 \sum_{j=1}^n p_j k_j x^*_j.
\end{equation}
The left-hand side equals $z$ whereas the right side equals $2z\Theta^*$, 
implying $\Theta^*=1/2$. Since, at equilibrium, we have
\begin{equation}
x^*_i=\frac{\lambda k_i\Theta^*}{1+\lambda k_i\Theta^*}=\frac{\lambda k_i}{2+\lambda k_i},
\end{equation}
from the definition of $\Theta^*$, we obtain the following relation for a 
local extremum:
\begin{equation}
z\Theta^*=\sum_{i=1}^n p_i k_i \frac{\lambda k_i}{2+\lambda k_i}=\frac{z}{2}.
\end{equation}
Together with the probability and the average degree constraints, a local 
extremum with $\bar{\psi}(\mathbf{p},\mathbf{p}^{\prime})=0$ can be found in 
the set $\mathcal{P}^o$ where any $p\in \mathcal{P}^o$ satisfies
\begin{equation}\label{eq:po_def}
\begin{split}
&\sum_{i=1}^n p_i \frac{\lambda k^2_i}{2+\lambda k_i}=\frac{z}{2},\qquad \sum_{i=1}^n p_i k_i=z,\\
&\sum_{i=1}^n p_i=1,\qquad p_i\in [0,1]\quad \forall\ i\in\{1,\cdots,n\}.
\end{split}
\end{equation}
The spreading prevalence for $\mathbf{p}\in \mathcal{P}^o$ can be directly 
obtained from the definition of $\mathcal{P}^o$, without solving the HMF 
equations. In particular, subtracting the second equation in 
Eq.~\eqref{eq:po_def} by the first equation on both sides, we have
\begin{equation}\label{eq:po_prevalence}
\sum_{i=1}^n p_i \frac{2 k_i}{2+\lambda k_i}=\frac{2}{\lambda}\sum_{i=1}^n p_i x_i=\frac{2}{\lambda}\psi(\mathbf{p})=\frac{z}{2},
\end{equation}
which implies $\psi(\mathbf{p})=\lambda z/4$ for $\mathbf{p}\in \mathcal{P}^o$.
That is, for any $\mathbf{p}\in \mathcal{P}^o$, the resulting spreading 
prevalence is the same.

For $\mathbf{p}\in \mathcal{P}^o$, conversely we have 
$\bar{\psi}(\mathbf{p},\mathbf{p}^{\prime})=0$ for all feasible directions. 
To see this, consider the definition of $\Theta^*$,
\begin{equation}\label{eq:Theta_star_def}
z\Theta^*=\sum_{i=1}^n p_i k_i \frac{\lambda k_i \Theta^*}{1+\lambda k_i \Theta^*}.
\end{equation}
If the right-hand side is viewed as a function of $\Theta^*$, then it 
increases with $\Theta^*$. For $\Theta^*=0$, the right-hand side equals zero 
and for $\Theta^*\to \infty$ it converges to $z$. Consequently, for a fixed 
$\mathbf{p}$, there is a unique $\Theta^*$ such that the right-hand side 
equals $z/2$. Since $\mathbf{p}\in \mathcal{P}^o$, from the first equation 
in Eq.~\eqref{eq:po_def}, we have $\Theta^*=1/2$ and then 
Eq.~\eqref{eq:local_label} holds. Similarly, for 
$\mathbf{p} \notin \mathcal{P}^o$, we have $\Theta^* \neq 1/2$. The conclusion
is that for $\mathbf{p}\in\mathcal{P}$, 
$\bar{\psi}(\mathbf{p},\mathbf{p}^{\prime})=0$ holds for all feasible 
directions if and only if $\mathbf{p}\in \mathcal{P}^o$.

\subsection{Necessary condition for the global optimal solution}\label{subsec:global}

Suppose $\mathcal{P}^o$ is nonempty, the question is as follows: Are the degree
distributions local maxima or a global maximum? As the set $\mathcal{P}^o$ is
defined through simple linear equations, we can prove that any
$\mathbf{p}\in \mathcal{P}^o$ is indeed a global maximum via 
algebraic manipulations. Concretely, in the following, we prove that
if $\mathbf{p}\notin \mathcal{P}^o$, then $\psi(\mathbf{p})< \lambda z/4$.
When $\mathcal{P}^o$ is empty, we show that the support of the optimal
degree distribution has no more than two distinct nodal degrees.

For any $\mathbf{p}\notin \mathcal{P}^o$,
this is trivially true if $\Theta^*=0$ and we assume $\Theta^*> 0$. Suppose 
there exists 
$\mathbf{p}\notin \mathcal{P}^o$ but $\psi(\mathbf{p})\geq \lambda z/4$, then 
from the definition of $\psi(\mathbf{p})$, we have
\begin{equation}
\frac{1}{\lambda \Theta^*}\psi(\mathbf{p})=\sum_{i=1}^n p_i \frac{k_i}{1+\lambda k_i \Theta^*}\geq \frac{z}{4 \Theta^*}.
\end{equation}
Subtracting $\sum_{i=1}^n p_i k_i=z$ from the inequality on both sides, we have
\begin{equation}
\sum_{i=1}^n p_i k_i \frac{\lambda k_i \Theta^*}{1+\lambda k_i \Theta^*}=z\Theta^*\leq z-\frac{z}{4 \Theta^*}.
\end{equation}
The inequality implies $(2\Theta^*-1)^2\leq 0$. An equality holds only when 
$\Theta^*=1/2$, but this contradicts with $\Theta^*\neq 1/2$ for 
$p\notin \mathcal{P}^o$ from the discussions below 
Eq.~\eqref{eq:Theta_star_def}.

The analysis so far reveals that, when $\mathcal{P}^o$ is nonempty, any 
$\mathbf{p}$ is a global maximum if and only if it belongs to $\mathcal{P}^o$. 
It remains to address the following issues. (\romannumeral1) For which values 
of $\lambda$ is the set $\mathcal{P}^o$ nonempty? (\romannumeral2) If 
$\mathcal{P}^o$ is empty, how do we find the local maxima with 
$\bar{\psi}(\mathbf{p},\mathbf{p}^{\prime})<0$ for all feasible perturbations. 
We will solve (\romannumeral2) partly for the rest of this section, and 
provide full answers to (\romannumeral1) and (\romannumeral2) in the next section.

Suppose $\mathcal{P}^o$ is empty. Consider any $\mathbf{p}\in\mathcal{P}$ and 
define the support of $\mathbf{p}$ as 
$\mathrm{supp}(\mathbf{p})=\{k_i:p_i> 0\}$. Suppose $\mathrm{supp}(\mathbf{p})$
has more than two distinct nodal degrees, we can pick any 
$m\geq 3$ points $\{k_{i_1},k_{i_2},\cdots k_{i_m}\}\subset\mathrm{supp}(\mathbf{p})$ 
from the support of $\mathbf{p}$ and consider a perturbation $\bar{\mathbf{p}}$
whose elements are nonzero only at these points. For any $\bar{\mathbf{p}}$ 
which is nonzero only on the support of $\mathbf{p}$, we can always choose 
$\alpha$ sufficiently small such that
\begin{equation}\label{eq:suppt_1}
\mathbf{p}+\alpha \bar{\mathbf{p}}\in [0\ 1]^n,\ \ \mathbf{p}-\alpha \bar{\mathbf{p}}\in [0\ 1]^n.
\end{equation}
The perturbations $\alpha \bar{\mathbf{p}}$ and $-\alpha\bar{\mathbf{p}}$ 
are thus both feasible for sufficiently small $\alpha$ . As $\mathcal{P}^o$ is 
empty, there always exists $\bar{\mathbf{p}}$ such that 
$\mathcal{Z}^{(m)}\bar{\mathbf{p}}\neq 0$. From Eq.~\eqref{eq:bpsi_2}, we have
\begin{equation}\label{eq:suppt_2}
\bar{\psi}\left(\mathbf{p},\mathbf{p}+\alpha \bar{\mathbf{p}}\right)=-\bar{\psi}\left(\mathbf{p},\mathbf{p}-\alpha \bar{\mathbf{p}}\right).
\end{equation}
This indicates that if $\mathcal{P}^o$ is empty, then any $\mathbf{p}$ whose 
support has more than two distinct degrees cannot be a local maximum and the 
optimal $\mathbf{p}^o$ must concentrate on no more than two distinct nodal 
degrees.

\section{Finding the optimal degree distributions} \label{sec:find_optim}

The results in Sec.~\ref{sec:necess_cond} indicate that, to find the optimal 
distributions, it is only necessary to determine whether set 
$\mathcal{P}^o$ is nonempty. If it is empty, the task is to search through 
all degree distributions whose support consists of one or two nodal degrees.
In fact, in the latter case, the HMF equation can be solved analytically to
yield the optimal degree distributions.

\subsection{Conditions for $\mathbf{\mathcal{P}^o}$ to be nonempty}
\label{subsec:nonempty}

As $\mathcal{P}^o$ is a closed convex set, by the Krein-Milman theorem, it 
is the convex hull of all its extremum points (i.e., $p\in \mathcal{P}^o$ that 
does not lie in the open line segment joining any two other points in 
$\mathcal{P}^o$). To check if $\mathcal{P}^o$ is nonempty is equivalent to 
examining if all its extremum points exist. In the following, we first show 
that the support of the extremum points of $\mathcal{P}^o$ has no more than three distinct nodal degrees. In this case, the value of $\mathbf{p}$ is uniquely determined by choice of
the support. As a result, we can solve $\mathbf{p}$ in terms of the support 
and $\lambda$ explicitly. With a fixed chosen support and the $\lambda$ value
so determined, the corresponding $\mathbf{p}$ is physical for 
$\mathbf{p}\in [0,1]^n$.
By checking all the points that are supported on no more than three degrees, 
we can derive the condition for $\lambda$ under which $\mathcal{P}^o$ is 
nonempty.

Suppose there exists $\mathbf{p}\in \mathcal{P}^o$ whose support has more 
than three degrees. Pick any $m\geq 4$ points 
$\{k_{i_1},k_{i_2},\cdots k_{i_m}\}\subset \mathrm{supp}(\mathbf{p})$ and 
consider a perturbation $\bar{\mathbf{p}}$ whose elements are nonzero only 
on these points. Define
\begin{equation}
\mathcal{Y}^{(m)}=\left(
\begin{array}{cccc}
1 & 1 & \cdots & 1\\
k_{i_1} & k_{i_2} & \cdots & k_{i_m}\\
\frac{\lambda k^2_{i_1}}{2+\lambda k_{i_1}} & \frac{\lambda k^2_{i_2}}{2+\lambda k_{i_2}} & \cdots & \frac{\lambda k^2_{i_m}}{2+\lambda k_{i_m}}
\end{array}\right).
\end{equation} 
A feasible direction of perturbation $\bar{\mathbf{p}}$, which keeps 
$\mathbf{p}\pm \alpha \bar{\mathbf{p}}$ staying inside $\mathcal{P}^o$ for 
sufficiently small values of $\alpha$, must satisfy the condition
$\mathcal{Y}^{(m)}\bar{\mathbf{p}}=0$. The nullity of $\mathcal{Y}^{(m)}$ is 
$\mathrm{nullity}(\mathcal{Y}^{(m)})=m-3$. Thus, for $m>3$, the space of 
feasible perturbations is nonempty. Moreover, we can always choose 
$\alpha_1>0$ and $\alpha_2>0$ such that the support of 
$\mathbf{p}+\alpha_1\bar{\mathbf{p}}$ and $\mathbf{p}-\alpha_2\bar{\mathbf{p}}$
has $m-1$ distinct nodal degrees. In this way, $\mathbf{p}$ lies on the open 
line segment that joins $\mathbf{p}+\alpha_1 \bar{\mathbf{p}}$ and 
$\mathbf{p}-\alpha_2 \bar{\mathbf{p}}$. This means that, if the support of 
$\mathbf{p}\in \mathcal{P}^o$ has more than three distinct nodal degrees, 
it will not be an extremum point of $\mathcal{P}^o$. 

To determine if $\mathcal{P}^o$ is nonempty, it thus suffices to check 
if there exists $\mathbf{p}\in \mathcal{P}^o$ whose support has no more than 
than three distinct nodal degrees. Consider any $k_{i_1}>k_{i_2}>k_{i_3}$, 
the values of $p_{i_1}$, $p_{i_2}$ and $p_{i_3}$ are uniquely determined by 
Eq.~\eqref{eq:po_def}, which are
\begin{equation}\label{eq:p3_formula}
\begin{split}
&p_{i_1}=-\frac{(k_{i_1}\lambda + 2)g(k_{i_2},k_{i_3})}{8 \lambda (k_{i_1} - k_{i_2})(k_{i_1} - k_{i_3})},\\
&p_{i_2}=+\frac{(k_{i_2} \lambda + 2) g(k_{i_1},k_{i_3})}{8 \lambda (k_{i_1} - k_{i_2})(k_{i_2} - k_{i_3})},\\
&p_{i_3}=-\frac{(k_{i_3}\lambda + 2)g(k_{i_1},k_{i_2})}{8 \lambda(k_{i_1} - k_{i_3})(k_{i_2} - k_{i_3})},
\end{split}
\end{equation}
where 
\begin{equation}\label{eq:g_def}
g(k_a,k_b)=\left(\lambda^2 z -4\lambda \right)k_a k_b+2\lambda z  (k_a+k_b)-4 z.
\end{equation}
The degree distribution is physically meaningful insofar as 
$p_{i_1},p_{i_2},p_{i_3}\in [0\ 1]$. Since $p_{i_1}+p_{i_2}+p_{i_3}=1$, it is 
sufficient to guarantee $p_{i_1},p_{i_2}$ and $p_{i_3}$ to be nonnegative, i.e.,
to guarantee
\begin{equation}\label{eq:po_cond}
g(k_{i_2},k_{i_3})\leq 0,\quad \ g(k_{i_1},k_{i_2})\leq 0,\quad  g(k_{i_1},k_{i_3})\geq 0.
\end{equation}
In Appendix~\ref{appendix_D}, we analyze the three inequalities in detail. 
Here we summarize the procedure and results. We study under what conditions 
the three inequalities in \eqref{eq:po_cond} hold consecutively. 
Particularly, we first derive the condition for the existence of 
$(k_{i_1},k_{i_3})$ such that $g(k_{i_1},k_{i_3})\geq 0$ holds. Then, under 
this condition, we check if there exists $k_{i_2}$ such that the other two 
inequalities in \eqref{eq:po_cond} hold. Consider the inequality 
$g(k_{i_1},k_{i_3})\geq 0$. The possible values of the two nodal degrees 
are $k_{i_1}\in \{k_{1},k_2,\cdots,z^{+}\}$ and 
$k_{i_3}\in\{z^{-},\cdots,k_{n-1},k_n\}$, where 
$z^{+}=\mathrm{min}_i \{k_i\geq z\}$ and $z^{-}=\mathrm{max}_i \{k_i\leq z\}$.
As $g(k_a,k_b)$ is quadratic in $\lambda$, we can show that $g(k_a,k_b)\geq 0$ 
if $\lambda\geq \lambda^{(k_a,k_b)}$ but $g(k_a,k_b)< 0$ otherwise, where
\begin{equation}\label{eq:lambda_ab}
\lambda^{(k_a,k_b)} =\frac{2}{z}-\frac{1}{k_a}-\frac{1}{k_b}+\sqrt{\left(\frac{1}{k_a}+\frac{1}{k_b}-\frac{2}{z}\right)^2+\frac{4}{k_a k_b}}. 
\end{equation}
As $\lambda^{(k_a,k_b)}$ is a decreasing function of $k_a$ for $k_a\geq z^+$ 
and an increasing function of $k_b$ for $k_b\leq z^-$, we can show that there 
exists $(k_{i_1},k_{i_3})$ such that $g(k_{i_1},k_{i_3})\geq 0$ holds insofar
as $\lambda \geq \lambda_2$, where $\lambda_2=\lambda^{(k_1,k_n)}$. 
Furthermore, when this condition holds, we can show that there exists 
$k_{i_2}$ such that the other two inequalities in \eqref{eq:po_cond} hold if 
and only if $\lambda\leq \lambda_3=\lambda^{(z^+,z^{-})}$.

Overall, the values of $\lambda$ are divided by $\lambda_1,\lambda_2$, and 
$\lambda_3$ into four regions, where $\lambda_1$ is defined in 
Eq.~\eqref{eq:lambda1_def}. The four regions are described as follows. \\
\indent
(\romannumeral1) For $\lambda\leq \lambda_1$, the optimization problem is 
trivial, i.e., no degree distribution can trigger an epidemic outbreak. \\
\indent
(\romannumeral2) For $\lambda_1<\lambda< \lambda_2$, set $\mathcal{P}^o$ 
is empty, thus the global maximum can only be found among all $\mathbf{p}$ 
supported on one or two nodal degrees. \\
\indent
(\romannumeral3) For $\lambda_2\leq \lambda \leq \lambda_3$, set 
$\mathcal{P}^o$ is nonempty and any $\mathbf{p}\in \mathcal{P}^o$ will lead 
to equal spreading prevalence $\lambda z/4$. In Appendix~\ref{appendix_D},
we show that for $\lambda=\lambda_2$, set $\mathcal{P}^o$ consists of a 
unique degree distribution supported on $\{k_1,k_n\}$, whereas for $\lambda=\lambda_3$, set $\mathcal{P}^o$ has a unique degree distribution 
supported on $\{z^+,z^-\}$. For $\lambda_2<\lambda<\lambda_3$, there are 
infinitely many global maxima that constitute a plateau  of equal spreading 
prevalence. \\
\indent
(\romannumeral4) For $\lambda>\lambda_3$, set $\mathcal{P}^o$ again 
becomes empty, and the global maxima can only be supported on one or two 
nodal degrees.

\subsection{Analytic solutions of HMF equations on one or two degrees}
\label{subsec:two_solution}

Having determined the conditions under which $\mathcal{P}^o$ is nonempty, we are now in a position to find the optimal degree distributions that are supported on one or two degrees. 
In this case, the HMF equations consist of only one or two different equations
so the equilibrium solution can be solved explicitly. We can then directly 
optimize the solution to obtain the optimal degree distribution on one or 
two nodal degrees.

Consider the situation where $\mathbf{p}$ is supported on one or two different 
nodal degrees. Let $k_1\geq k_{i_1}\geq z^+$ and $z^-\geq k_{i_2}\geq k_n$ be 
any two nodal degrees from $\mathbf{k}$ so that $p_{i_1}$ and $p_{i_2}$ are 
uniquely determined by 
\begin{equation}
p_{i_1}+p_{i_2}=1,\qquad p_{i_1} k_{i_1}+p_{i_2} k_{i_2}=z,
\end{equation}
which leads to the solutions of $p_{i_1}$ and $p_{i_2}$ in terms of $k_{i_1},k_{i_2}$, and $z$ as
\begin{equation} \label{eq:p_ab}
p_{i_1}=\frac{z-k_{i_2}}{k_{i_1}-k_{i_2}},\qquad p_{i_2}=\frac{k_{i_1}-z}{k_{i_1}-k_{i_2}}.
\end{equation}
When $z$ is an integer and either $k_{i_1}$ or $k_{i_2}$ equals $z$, it reduces
to the case where $\mathbf{p}$ is supported on one nodal degree. With the 
values of $p_{i_1}$ and $p_{i_2}$, the HFM equation can be solved analytically 
(Appendix~\ref{appendix_E}). After some algebraic manipulations, we obtain
the spreading prevalence as
\begin{equation}\label{eq:psi_two}
\begin{split}
\psi(\mathbf{p})=&1-u-\frac{1}{\lambda z} \left( u^2+ v^2\right)
\\
&+\frac{u}{\lambda z} \sqrt{\lambda z \left(\lambda z - 4  +4u\right) +  4 v^2}.
\end{split}
\end{equation} 
where
\begin{equation}\label{eq:uv_def}
u=\frac{1}{2}\left(\frac{z}{k_{i_1}}+\frac{z}{k_{i_2}}\right),\qquad v=\frac{1}{2}\left(\frac{z}{k_{i_1}}-\frac{z}{k_{i_2}}\right).
\end{equation}
The degrees $k_{i_1}$ and $k_{i_2}$ are then uniquely determined by the values 
of $u$ and $v$. 

We can now carry out optimization among all degree distributions that are 
supported on one or two nodal degrees. The goal is to find the optimal 
degree values $k_{i_1}$ and $k_{i_2}$ such that $\psi(\mathbf{p})$ given by 
Eq.~\eqref{eq:psi_two} is maximized. Our approach is to treat $k_{i_1}$ and 
$k_{i_2}$ as continuous variables to obtain the maxima of $\psi(\mathbf{p})$, 
which can finally be used to find the actual optimal values of $k_{i_1}$ and 
$k_{i_2}$ as integers.

From Eq.~\eqref{eq:p_ab}, we see that $p_{i_1}$ and $p_{i_2}$ are uniquely 
determined by the choice of $k_{i_1}$ and $k_{i_2}$ which, in turn, are 
uniquely determined by the values of $u$ and $v$ defined in 
Eq.~\eqref{eq:uv_def}. The equivalent problem is to optimize $\psi(\mathbf{p})$
by $u$ and $v$. It is convenient to rewrite $\psi(\mathbf{p})$ as $\psi(u,v)$. 
Taking the partial derivatives of $\psi(u,v)$, we obtain
\begin{eqnarray} 
\nonumber	
\frac{\partial \psi(u,v)}{\partial u} &=&\left(\frac{1}{\lambda z} \sqrt{\lambda z \left(\lambda z - 4  +4u\right) + 4v^2}-1\right) \\ \label{eq:u_derivative}
	&& \times \left(1-\frac{2u}{\sqrt{\lambda z \left(\lambda z - 4  +4u\right) + 4 v^2}}\right), \\ \label{eq:v_derivative}
	\frac{\partial \psi(u,v)}{\partial v}&=&\frac{2v}{\lambda z}\left(\frac{2u}{\sqrt{\lambda z \left(\lambda z - 4  +4u\right) + 4 v^2}}-1\right).
\end{eqnarray}
The two partial derivatives vanish simultaneously only for
\begin{equation} \label{eq:uv_relation}
2u=\sqrt{\lambda z \left(\lambda z - 4  +4u\right) + 4 v^2},
\end{equation}
which defines a curve on the $u\hbox{-}v$ plane where every point on it is 
a critical point of $\psi(u,v)$. Substituting Eq.~\eqref{eq:uv_relation} into 
Eq.~\eqref{eq:psi_two}, we obtain the spreading prevalence along the curve as 
\begin{equation}\label{eq:optimal_prevalence}
	\psi(\mathbf{p})=\frac{\lambda z}{4},
\end{equation}
which is exactly the spreading prevalence for those 
$\mathbf{p}\in \mathcal{P}^o$, given that $\mathcal{P}^o$ is nonempty.

Substituting the definition of $u$ and $v$ in Eq.~\eqref{eq:uv_def} into 
Eq.~\eqref{eq:uv_relation}, we can express the curve in terms of $k_{i_1}$ 
and $k_{i_1}$ as $g(k_{i_1},k_{i_2})=0$, where
\begin{equation}
g(k_a,k_b)=(\lambda^2z -4\lambda)k_a k_b +2\lambda z (k_a +k_b)  -4 z.
\end{equation}
This function is also exactly the same as Eq.~\eqref{eq:g_def}, the one that 
emerges when we analyze the extremum points of $\mathcal{P}^o$. Not all points 
$(k_a,k_b)$ along the optimal curve in Eq.~\eqref{eq:uv_relation} are 
physically meaningful. Especially, for a point on the $k_a\hbox{-}k_b$ plane 
to be meaningful, it must be an integer point that lies in the region,
\begin{equation}
R=\{(k_a,k_b): k_1\geq k_a\geq z^+,\ z^{-}\geq k_b\geq k_n \}.
\end{equation}
From the discussions below Eq.~\eqref{eq:g_def}, the curve $g(k_a,k_b)=0$ 
passes an integer point $(k_a,k_b)$ when $\lambda=\lambda^{(k_a,k_b)}$, 
where $\lambda^{(k_a,k_b)}$ is defined in Eq.~\eqref{eq:lambda_ab}. When 
this happens, the degree distribution supported on $\{k_a,k_b\}$ belongs to set $\mathcal{P}^o$. In fact, if we let $(k_{i_1},k_{i_3})=(k_a,k_b)$ and 
substitute $g(k_{i_1},k_{i_3})=0$ into Eq.~\eqref{eq:p3_formula}, we then 
have $p_{i_2}=0$ and 
\begin{equation}
p_{i_1}=\frac{z-k_{i_3}}{k_{i_1}-k_{i_3}},\qquad p_{i_3}=\frac{k_{i_1}-z}{k_{i_1}-k_{i_3}}.
\end{equation}
This recovers exactly the same degree distribution defined in 
Eq.~\eqref{eq:p_ab}. For $\lambda<\lambda_2$ or $\lambda>\lambda_3$, set 
$\mathcal{P}^o$ is empty, and no integer point in region $R$ can lie on 
the curve $g(k_a,k_b)=0$. In this case, it is necessary to further analyze 
the optimal degree distribution.

For convenience, we write $\psi(\mathbf{p})$ as $\psi(k_{i_1},k_{i_2})$ and 
have
\begin{equation} \label{eq:psi_k}
\frac{\partial \psi(k_{i_1},k_{i_2})}{\partial k_{i_1}}=-\frac{z}{2k_{i_1}^2}\left(\frac{\partial \psi(u,v)}{\partial u}+\frac{\partial \psi(u,v)}{\partial v}\right).
\end{equation}
Substituting Eqs.~\eqref{eq:u_derivative} and \eqref{eq:v_derivative} into
Eq.~\eqref{eq:psi_k}, we get
\begin{equation} \label{eq:d_psi_ka}
\begin{split}
&\frac{\partial \psi(k_{i_1},k_{i_2})}{\partial k_{i_1}}=\left(1-\frac{2u}{\sqrt{\lambda z \left(\lambda z - 4  +4u\right) + 4 v^2}}\right)\\
& \times \frac{z}{2k_{i_1}^2}\left(\frac{2v}{\lambda z}+1-\frac{1}{\lambda z} \sqrt{\lambda z \left(\lambda z - 4  +4u\right) + 4v^2}\right).
\end{split}
\end{equation}
Since $u-v=z/k_{i_2}>1$, we have
\begin{equation}
\begin{split}
&\sqrt{\lambda z \left(\lambda z - 4  +4u\right) + 4v^2}\\
>&\sqrt{\lambda z \left(\lambda z +4v\right) + 4v^2}>2v+\lambda z.
\end{split}
\end{equation} 
The last line in Eq.~\eqref{eq:d_psi_ka} is, thus, negative. For
\begin{equation}\label{eq:posi_cond}
\sqrt{\lambda z \left(\lambda z - 4  +4u\right) + 4 v^2}-2u>0,
\end{equation}
$\psi(k_{i_1},k_{i_2})$ is a decreasing function of $k_{i_2}$; otherwise it 
is an increasing function of $k_{i_1}$. Similarly, the partial derivative of 
$\psi(k_{i_1},k_{i_2})$ with respect to $k_{i_2}$ is
\begin{equation}
\begin{split}
&\frac{\partial \psi (k_{i_1},k_{i_2})}{\partial k_{i_2}}=\left(1-\frac{2u}{\sqrt{\lambda z \left(\lambda z - 4  +4u\right) + 4 v^2}}\right)\\
& \times \frac{z}{2k_{i_2}^2}\left(-\frac{2v}{\lambda z}+1-\frac{1}{\lambda z} \sqrt{\lambda z \left(\lambda z - 4  +4u\right)+ 4v^2} \right),
\end{split}
\end{equation}
where the term in the last line is positive. Thus, if Eq.~\eqref{eq:posi_cond} 
holds, $\psi(k_{i_1},k_{i_2})$ is an increasing function of $k_{i_2}$, 
otherwise it is a decreasing function of $k_{i_2}$.

Recall that $g(k_a,k_b)$ in Eq.~\eqref{eq:g_def} is equivalent to the relation 
in Eq.~\eqref{eq:uv_relation}. The inequality in Eq.~\eqref{eq:posi_cond} can 
then be written in terms of $k_{i_1}$ and $k_{i_2}$ as
\begin{equation}
g(k_{i_1},k_{i_2})>0.
\end{equation}
From the discussions in Appendix~\ref{appendix_D}, for any $(k_a,k_b)\in R$, 
we have $g(k_a,k_b)<0$ if $\lambda<\lambda_2$ and $g(k_a,k_b)>0$ if 
$\lambda>\lambda_3$. These results lead to the optimal degree distributions 
in each of the parameter regions of $\lambda$.

For $\lambda_1<\lambda<\lambda_2$, we have $g(k_a,k_b)<0$ for any 
$(k_a,k_b)\in R$. Consequently, $\psi(k_{i_1},k_{i_2})$ is an increasing 
function of $k_{i_1}$ and a decreasing function of $k_{i_2}$. In this case, 
the optimal degree distribution is supported on $k_a=k_1$ and $k_b=k_n$. 
Moreover, the spreading prevalence of the optimal degree distribution is 
strictly less than $\lambda z/4$.

For $\lambda_2\leq \lambda\leq \lambda_3$, the degree distribution 
$\mathbf{p}$ is a global maximum if and only if $\mathbf{p}\in \mathcal{P}^o$. 
Since $\mathcal{P}^o$ is a connected set, all the global maxima constitute a 
plateau of degree distributions with equal spreading prevalence. For 
$\lambda=\lambda_2$, the set $\mathcal{P}^o$ consists of a unique degree 
distribution, which is exactly the optimal one for $\lambda<\lambda_2$. For 
$\lambda=\lambda_3$, the set $\mathcal{P}^o$ also has one unique degree 
distribution, and we will see that it is the optimal one for 
$\lambda>\lambda_3$.

For $\lambda>\lambda_3$, we have $g(k_a,k_b)>0$ for any $(k_a,k_b)\in R$. 
As a result, $\psi(k_{i_1},k_{i_2})$ is a decreasing function of $k_{i_1}$ 
and an increasing function of $k_{i_2}$. In this case, the optimal degree 
distribution is supported on $k_{i_1}=z^+$ and $k_{i_2}=z^-$.

\section{Characteristics of optimal degree distributions} \label{sec:properties}

For relatively low infection rates ($\lambda_1 < \lambda \leq \lambda_2$), 
the optimal degree distribution is supported on the maximal and minima possible
degrees $\{k_1,k_n\}$. For high infection rates ($\lambda\geq \lambda_3$), the 
optimal degree distribution is supported on the two nodal degrees $\{z^+,z^-\}$
that are nearest to the average degree $z$. Therefore, we need to study how the
support of the optimal degree distributions behaves for intermediate infection 
rates in the range $[\lambda_2,\lambda_3]$. Let 
$\mathcal{P}^{e}\subset \mathcal{P}^o$ be the set of all extremum points of 
$\mathcal{P}^o$, where $\mathcal{P}^e$ is a finite set. As $\mathcal{P}^o$ is 
the convex hull of all its extremum points, for any 
$\mathbf{p}\in \mathcal{P}^o$, it is a convex combination of the extremum points,
\begin{equation}\label{eq:convex_comb}
\mathbf{p}=\sum_{\mathbf{p}^{e}\in\mathcal{P}^e} c\left(\mathbf{p}^{e}\right) \mathbf{p}^{e},
\end{equation}
where $c\left(\mathbf{p}^{e}\right)\geq 0$ and 
$\sum_{\mathbf{p}^{e}\in\mathcal{P}^e} c\left(\mathbf{p}^{e}\right)=1$. 
The broadest support (i.e., the support with the largest number of distinct 
nodal degrees) of $\mathbf{p}\in \mathcal{P}^o$ thus is 
\begin{equation}
\bigcup_{\mathbf{p}^{e}\in\mathcal{P}^e} \mathrm{supp}\left(\mathbf{p}^{e}\right).
\end{equation}
Any degree distribution with $c\left(\mathbf{p}^{e}\right)> 0$ for all 
$\mathbf{p}^{e}\in\mathcal{P}^e$ will have the broadest possible support. 

\begin{figure} [ht!]
\centering
\includegraphics[width=0.8\linewidth]{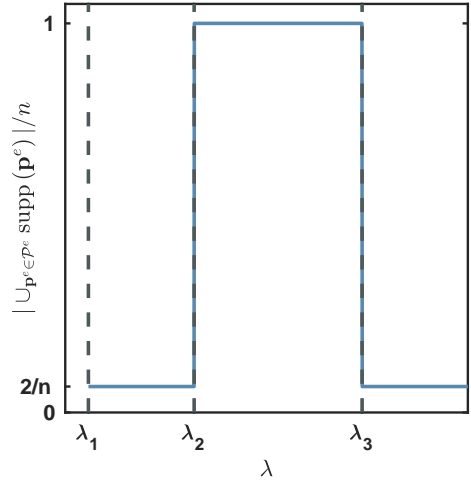}
\caption{Normalized cardinality of the broadest support for 
$\mathbf{p}\in \mathcal{P}^o$ versus  $\lambda$. The vertical gray dashed 
lines mark the locations of $\lambda_1, \lambda_2$, and $\lambda_3$ that 
divide the values of $\lambda$ into different regions. For 
$\lambda\leq \lambda_2$ or $\lambda\geq \lambda_3$, the normalized cardinality 
is $2/n$, whereas it is one for $\lambda_2<\lambda<\lambda_3$. 
For $\lambda_2<\lambda<\lambda_3$, the cardinality of the broadest possible 
support is obtained by testing all the extremum points numerically. The values
of other parameters are $k_1=30, k_n=1$, and $z=15.5$. 
The values of $\lambda_i$ for $i\in \{1\hbox{--}3\}$ are $\lambda_1\approx 0.0344$, 
$\lambda_2\approx 0.0709$ and $\lambda_3\approx 0.1290$. The support of the 
degree distribution can take on any integer value between $k_1$ and $k_n$, 
i.e., $\mathbf{k}=[30,29,\cdots,1]^T$ and $n=30$.}
\label{fig:support}
\end{figure}

Consider the case where $\lambda$ is slightly above $\lambda_2$ and $(k_1,k_n)$
is a unique point such that $g(k_1,k_n)>0$. From the discussions at the end of 
Appendix~\ref{appendix_D}, we have that, by choosing $k_{i_1}=k_1, k_{i_3}=k_n$, and $k_{i_2}$ to be any allowed degree with $k_1>k_{i_2}>k_3$, 
the triple $(k_{i_1},k_{i_2},k_{i_3})$ will define a physical degree 
distribution from Eq.~\eqref{eq:p3_formula}. As the middle degree $k_{i_2}$ 
is arbitrary, the broadest support in this case consists of all the allowed 
degrees in $\mathbf{k}$, i.e., the cardinality of the broadest support 
increases abruptly from 2 to $n$ at $\lambda=\lambda_2$. Similarly, it can 
be seen that, when $\lambda$ is slightly below $\lambda_3$ and $(z^+,z^-)$ is 
the unique point such that $g(z^+,g^-)<0$, the broadest support also consists 
of all the possible nodal degrees. Figure~\ref{fig:support} shows the 
normalized cardinality of the broadest possible support versus $\lambda$. 
We see that, for $\lambda_2<\lambda<\lambda_3$, the broadest support indeed
consists of all the distinct degrees allowed in $\mathbf{k}$, indicating that, 
except for relatively low or high values of $\lambda$, the support of the 
optimal degree distribution can be quite broad.

\begin{figure} [ht!]
\centering
\includegraphics[width=0.8\linewidth]{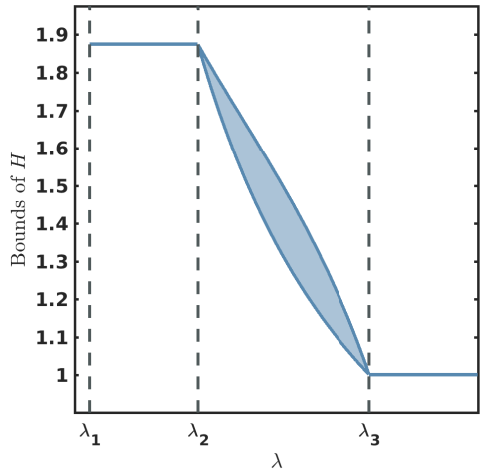}
\caption{Bounds of degree heterogeneity of the optimal degree distributions 
versus the infection rate. The vertical gray dashed lines mark the locations 
of $\lambda_1, \lambda_2$, and $\lambda_3$ that divide $\lambda$ into 
different regions. The blue solid trace represents the bounds of the 
degree heterogeneity $H$. For $\lambda\leq \lambda_2$ or 
$\lambda\geq \lambda_3$, the lower and upper bounds coincide. For 
$\lambda_2<\lambda<\lambda_3$, the degree heterogeneity can take on any value 
in the shaded region. Other parameter values are $k_1=30$, $k_n=1$ and 
$z=15.5$. The values of $\lambda_i$ for $i\in \{1\hbox{--}3\}$ and $n$ 
are the same as those in Fig.~\ref{fig:support}.}
\label{fig:H}
\end{figure}

In general, the degree heterogeneity of a network, defined as 
$H=\langle k^2\rangle/\langle k\rangle^2$, can have significant impacts on the 
spreading dynamics. A natural question is, what is the degree heterogeneity of 
the optimal degree distribution? Since the average degree is fixed 
($\langle k\rangle=z$), the degree heterogeneity determines the outbreak 
threshold. For sufficiently small values of $\lambda$ where there is a unique 
network that can trigger an epidemic outbreak, the optimal network structure 
is one with the largest degree heterogeneity. 

Consider the general problem of finding maxima and minima of $H$ among 
all degree distributions. The extrema of $H$ can be found by maximizing or 
minimizing the second moment $\langle k^2\rangle$ of the degree distribution. 
The Bhatia-Davis inequality stipulates that the second moment of $\mathbf{p}$ 
is maximized when it is concentrated at the endpoints $k_1$ and $k_n$. To 
minimize the second moment, we note that the definition 
$\langle k^2\rangle=\sum_{i=1}^n p_i k_i^2$ has a similar form to 
Eq.~\eqref{eq:bpsi_2} with $\chi_i$ replaced by $k_i^2$. Following the 
reasoning in Sec.~\ref{subsec:consequences}, we see that the minimum of $H$ 
is supported on two nodal degrees. Through a direct comparison of all 
distributions supported on two degrees, we find that $H$ is minimized when 
$\mathbf{p}$ concentrates on $\{z^+,z^-\}$. We see that the optimal degree 
distributions for $\lambda\leq \lambda_2$ and $\lambda\geq \lambda_3$ are 
exactly the ones that maximize and minimize the degree heterogeneity, 
respectively.

For a fixed $\lambda$ value in the intermediate region 
($\lambda_2<\lambda<\lambda_3$), the values of $H$ for different degree 
distributions in $\mathcal{P}^o$ are not necessarily identical. From 
Eq.~\eqref{eq:convex_comb}, we see that, if $\mathbf{p}$ is a convex 
combination of the extremum points, its second moment can be obtained by the 
same convex combination of the second moment of the extremum points. 
Consequently, the degree heterogeneity of $\mathbf{p}\in \mathcal{P}^o$ is 
bounded by that of the extremum points. Figure~\ref{fig:H} shows the bounds 
of the degree heterogeneity $H$ of the optimal degree distributions versus 
$\lambda$. The general phenomenon is that the optimal network is more 
heterogeneous for small infection rates but less so for large rates, as the 
upper and lower bound of $H$ decreases with $\lambda$. However, the degree 
heterogeneity does not decrease with $\lambda$ in a strict sense but only 
trendwise. In fact, if we draw a line segment joining the two degree 
distributions that reach the upper and lower bounds, then $H$ varies 
continuously on this line segment, i.e., the degree heterogeneity can take 
on any value between the lower and upper bounds. 

\begin{figure} [ht!]
\centering
\includegraphics[width=0.8\linewidth]{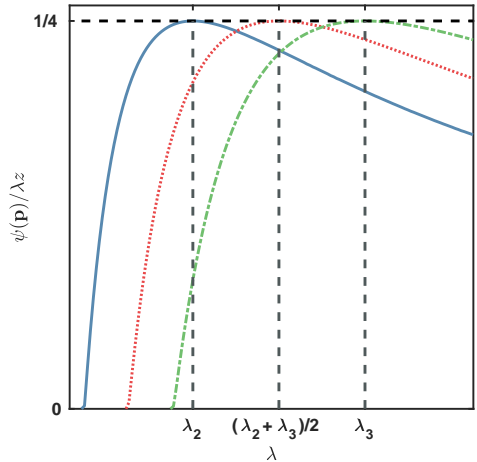}
\caption{Spreading prevalence divided by $\lambda z$ versus $\lambda$ for 
three different degree distributions. The values of the spreading prevalence 
are obtained by solving the HMF equations numerically. The vertical gray 
dashed lines mark the locations of $\lambda_2, (\lambda_2+\lambda_3)/2$, and $\lambda_3$ where the three degree distributions are optimal. The 
horizontal black dashed line correspond to $\psi(\mathbf{p})/\lambda z=1/4$. 
Other parameter values are $k_1=30, k_n=1$, and $z=15.5$. 
The values of $\lambda_i$ for $i\in \{1\hbox{--}3\}$ and $n$
are the same as those in Fig.~\ref{fig:support}.}
\label{fig:prevalence}
\end{figure}

Our analysis of the characteristics of the optimal degree distributions reveals
a phenomenon: The existence of a particular value of the infection rate for 
which every degree distribution is optimal. From the definition of 
$\mathcal{P}^o$, any $\mathbf{p}\in \mathcal{P}^o$ must satisfy the first 
equation in Eq.~\eqref{eq:po_def}, whose left-hand side is an increasing 
function of $\lambda$ that converges to zero or $z$ for $\lambda\to 0$ or
$\lambda \to \infty$, respectively. As a result, for any 
$\mathbf{p}\in \mathcal{P}$, there always exists a unique $\lambda$ value 
such that $\mathbf{p}\in \mathcal{P}^o$. Only two degree distributions are 
optimal under multiple values of $\lambda$, which are the two supported 
on either $\{k_1,k_n\}$ or $\{z^+,z^-\}$, as they are optimal when 
$\mathcal{P}^o$ is empty. In Sec.~\ref{subsec:consequences}, we have shown 
that for any $\mathbf{p}\notin \mathcal{P}^o$, its spreading prevalence is 
strictly less than $\lambda z/4$. This suggests the following phenomenon: 
For any degree distributions, its spreading prevalence as a function of 
$\lambda$ will touch the line $\psi(\mathbf{p})=\lambda z/4$ only at one 
value of $\lambda$ and under this value of $\lambda$ the degree distribution 
is among the optimal degree distributions. For all other values of $\lambda$, 
its spreading prevalence will strictly be below the line 
$\psi(\mathbf{p})=\lambda z/4$.

To illustrate the phenomenon, we consider three degree distributions that are 
optimal at $\lambda=\lambda_2, \lambda=(\lambda_2+\lambda_3)/2$, and 
$\lambda=\lambda_3$, respectively. For $\lambda=\lambda_2$ or 
$\lambda=\lambda_3$, the optimal degree distribution is unique. For 
$\lambda=(\lambda_2+\lambda_3)/2$, we randomly pick a degree distribution 
from $\mathcal{P}^o$ by a uniformly random convex combination of the extremum 
points. We plot $\psi(\mathbf{p})/\lambda z$ versus $\lambda$ for three 
degree distributions as shown in Fig.~\ref{fig:prevalence}. It can be seen
that the value of $\psi(\mathbf{p})/\lambda z$ reaches 1/4 at the predicted 
value of $\lambda$ and is below 1/4 for any other values of $\lambda$. 

\section{Discussion}\label{sec:discussion}

Given a dynamical process of interest, identifying the extremum network 
provides deeper insights into the interplay between network structure and 
dynamics. From the perspective of applications, searching for a global dynamics-specific optimal network can be valuable in areas such as 
information diffusion, transportation, and behavior promotion. The issue,
however, belongs to the category of dynamics-based inverse problems that
are generally challenging and extremely difficult to solve. We have taken 
an initial step in this direction. Specifically, by limiting the study to 
SIS type of spreading dynamics and imposing the annealed approximation, we 
have obtained analytic solutions to the inverse problem. Our solutions unveil 
a phenomenon with implications: A fundamental characteristic of the optimal 
network, its degree heterogeneity, depends on the infection rate. In 
particular, strong degree heterogeneity facilitates the spreading but only 
for small infection rates. For relatively large infection rates, the optimal 
structure tends to choose the networks that are less heterogeneous. This means 
that, when designing an optimal network, e.g., for information spreading, the 
ease with which information can diffuse among the nodes must be taken into 
account. Our analysis has also revealed the existence of a particular value 
of the infection rate for which every degree distribution is globally optimal.

The annealed approximation that serves the base of our analysis is applicable
to networks that are describable by the uncorrelated configuration model. It 
remains to be an open problem to find the optimal quenched networks for SIS 
dynamics. In Ref.~\cite{st2018phase}, the authors introduced a technique 
to bridge the annealed and quenched limit of the SIS model. The technique can 
provide a starting point to extend our analytic approach to quenched networks.
The variational analysis in the current paper can be extended to SIS type dynamics
on quenched weighted networks to derive a necessary condition for local optimum. 
In the variational calculus, we have to perform a network structural perturbation
to the mean-field equation; therefore, we emphasize a necessary element that makes
the variational calculations viable: The spreading prevalence is a continuous function
of the perturbations, at least locally around the network being perturbed. The variational analysis will result in a necessary condition for local optimum. However, it is
not clear yet what we can derive from the necessary condition without annealed
approximations. To generalize the theory to settings under less stringent 
simplifications is at present an open topic worth investigating.
Another assumption in the present paper is that only the number of 
edges is held fixed, and it is useful to study the optimal networks under 
more realistic restrictions. Moreover, it is of general interest to seek 
optimal solutions of network structures for different types of dynamical 
processes. Our paper represents a step forward in this direction.

\section*{Acknowledgments}

L.P. would like to acknowledge support from the National Natural Science Foundation 
of China under Grant No.~62006122. W.W. would like to acknowledge support from  the
National Natural Science Foundation of China under Grant No.~61903266, 
Sichuan Science and Technology Program Grant No.~20YYJC4001, China 
Postdoctoral Science Special Foundation Grant No.~2019T120829, and the Fundamental 
Research Funds for the central Universities Grant No.~YJ201830. L.T. would like to 
acknowledge support from The Major Program of the National Natural Science 
Foundation of China under Grant No.~71690242 and the National Key Research 
and Development Program of China under Grant No.~2020YFA0608601. Y.-C.L. would 
like to acknowledge support from the Vannevar Bush Faculty Fellowship program 
sponsored by the Basic Research Office of the Assistant Secretary of Defense 
for Research and Engineering and funded by the Office of Naval Research 
through Grant No.~N00014-16-1-2828.

\appendix

\section{Proof that $\mathbf{x}^*$ is a continuously differentiable function of $\mathbf{p}$ above the outbreak threshold} \label{appendix_A}

Setting the right-hand side of Eq.~\eqref{eq:HMF} to zero for an equilibrium, 
we get 
\begin{equation}\label{eq:appendix_A1}
x^*_i=\frac{\lambda k_i \Theta^*}{1+\lambda k_i \Theta^*}, \end{equation}
where $\Theta^*$ is obtained by substituting $\mathbf{x}(t)=\mathbf{x}^*$ into 
Eq.~\eqref{eq:Theta}. Define a function 
$f(\mathbf{p},\mathbf{x}):(\mathbf{p},\mathbf{x})\to \mathbb{R}^n$ as 
\begin{equation}
f_i(\mathbf{p},\mathbf{x})=x_i-\frac{\lambda k_i \sum_{l=1}^n p_l k_l x_l}{z+\lambda k_i \sum_{l=1}^n p_l k_l x_l},
\end{equation}
we have $f(\mathbf{p},\mathbf{x}^*)=0$. Note that $f(\mathbf{p},\mathbf{x})$ 
is a continuously differentiable function of $\mathbf{p}$ and $\mathbf{x}$. 
We now show that, from the relation $f(\mathbf{p},\mathbf{x}^*)=0$, the stable 
equilibrium point $\mathbf{x}^*$ can be written as a continuously 
differentiable function of $\mathbf{p}$ for $\lambda > z/\langle k^2\rangle$ 
by applying the implicit function theorem.

The derivative of $f_i(\mathbf{p},\mathbf{x})$ with respect to $x_j$ is 
\begin{equation}
\frac{\partial f_i(\mathbf{p},\mathbf{x})}{\partial x_j}=\delta_{i,j}-\frac{\lambda z k_i k_jp_j }{\left(z+\lambda k_i \sum_{l=1}^n p_l k_l x_l\right)^2},
\end{equation}
where $\delta_{i,j}$ is the Kronecker $\delta$. The Jacobian matrix of 
$f(\mathbf{p},\mathbf{x})$ to $\mathbf{x}$ can be written as 
$\mathcal{I}-\mathbf{r}\mathbf{s}^T$, where $\mathcal{I}$ is the $n\times n$ 
identity matrix and $\mathbf{r}$ and $\mathbf{s}$ are $n\times 1$ vectors with 
elements,
\begin{equation}
r_i=\frac{\lambda z k_i}{\left(z+\lambda k_i \sum_{l=1}^n p_l k_l x_l\right)^2},\qquad s_i=k_ip_i. 
\end{equation} 
Let $\mathbf{b}$ be an eigenvector of the matrix $\mathbf{r}\mathbf{s}^T$ 
with eigenvalue $\omega$. From the eigenvalue equation, we have
\begin{equation}
\sum_{j=1}^n r_i s_jb_j=\omega b_i.
\end{equation}
Multiplying both sides by $s_i$ and summing over $i$, we have
\begin{equation}
\sum_{i=1}^n r_i s_i \sum_{j=1}^n s_jb_j=\omega \sum_{i=1}^n s_i b_i.
\end{equation}
As a result, the only possible eigenvalues of matrix 
$\mathbf{r}\mathbf{s}^T$ are $\omega=0$ or $\omega=\sum_{i=1}^n r_i s_i$.

For $\lambda >z/\langle k^2\rangle$, all elements of $\mathbf{x}^*$ 
are positive. At $\mathbf{x}=\mathbf{x}^*$, we have
\begin{equation}
\begin{split}
\sum_{i=1}^n r_i s_i=&\sum_{i=1}^n \frac{\lambda z  p_i k_i^2 }{\left(z+\lambda k_i \sum_{l=1}^n p_l k_l x^*_l\right)^2}\\
=&\frac{1}{z}\sum_{i=1}^n \lambda p_i k_i^2 (1-x^*_i)^2\\
=&\frac{1}{z\Theta^*}\sum_{i=1}^n p_i k_i x^*_i (1-x^*_i)\\
=&1-\frac{1}{z\Theta^*}\sum_{i=1}^n p_i k_i \left(x^*_i\right)^2< 1,
\end{split}
\end{equation}
where the second and third equalities can be verified by substituting them
into Eq.~\eqref{eq:appendix_A1} and $x^*_i=\lambda k_i (1-x^*_i)\Theta^*$,
respectively.

Taken together, the eigenvalues of the matrix $\mathbf{r}\mathbf{s}^T$ are 
less than one for $\lambda > z/\langle k^2\rangle$, so the eigenvalues of 
the Jacobian matrix $\mathcal{I}-\mathbf{r}\mathbf{s}^T$ are less than zero, 
which further implies that the Jacobian matrix is invertible. By the implicit 
function theorem, $\mathbf{x}^*$ is a continuously differentiable function 
of $\mathbf{p}$.

\section{Eigenvalues of the Jacobian matrix} \label{appendix_B}

Denote the right side of Eq.~\eqref{eq:HMF} by
\begin{equation}
h_i=-x_i(t)+\lambda k_i \left[1-x_i(t)\right]\Theta.
\end{equation}
The Jacobian matrix for $h=\left(h_1,\cdots,h_n\right)^T$ at 
$\mathbf{x}=\mathbf{x}^*$ is exactly $\mathcal{J}$: $\nabla h=\mathcal{J}$. 
As $\mathbf{x}^*$ is the unique global stable equilibrium 
point~\cite{wang2008global}, the eigenvalues of $\mathcal{J}$ must have 
negative real parts.

\section{Detailed derivation of $\mathbf{\chi}$} \label{appendix_C}

Define two vectors $\mu$ and $\nu$ of length $n$ whose elements are
\begin{equation}
\mu_i= \frac{\lambda}{z} k_i\left(1-x^*_i\right),\qquad \nu_i=k_i p_i.
\end{equation}
Further, define a $n\times n$ diagonal matrix $\mathcal{D}$ with the elements
\begin{equation}
D_{ii}=-1-\lambda k_i\Theta ^*.
\end{equation}
By the Sherman-Morrison formula, we have
\begin{equation} \label{eq:appendix_B_local}
\begin{split}
&\mathcal{J}^{-1}=\left(\mathcal{D}+\mu\cdot\nu^T\right)^{-1}=\mathcal{D}^{-1}-\frac{\mathcal{D}^{-1}\cdot\mu\cdot\nu^T\cdot\mathcal{D}^{-1}}{1+\nu^T\cdot\mathcal{D}^{-1}\mu}.
\end{split}
\end{equation}
Substituting Eq.~\eqref{eq:bx_star} into Eq.~\eqref{eq:bpsi}, we get 
Eq.~\eqref{eq:bpsi_2} with $\chi_i$ given by
\begin{equation}\label{eq:chi_0}
\chi_{i}=x^*_i- k_i x^*_i p^T \mathcal{J}^{-1}\mu.
\end{equation}
Inserting Eq.~\eqref{eq:appendix_B_local} into Eq.~\eqref{eq:chi_0} leads to
\begin{equation}\label{eq:chi_B}
\begin{split}
&\chi_i=x^*_i \\
&+\frac{\lambda k_i x_i \sum_{j=1}^n p_j k_j(1-x^*_j)\left(1+\lambda k_j \Theta^*\right)^{-1}}{z-\lambda \sum_{j=1}^n p_j k^2_j(1-x^*_j)\left(1+\lambda k_j \Theta^*\right)^{-1}}.
\end{split}
\end{equation}
At the equilibrium point, we have
\begin{equation} \label{eq:B6}
- x^*_i+\lambda k_i\left(1-x^*_i\right)\Theta^*=0,
\end{equation}
which leads to 
\begin{equation}
\left(1+\lambda k_j \Theta^*\right)^{-1}=\left(1-x^*_i\right).
\end{equation}
Substituting the above two equations into Eq.~\eqref{eq:chi_B}, we obtain 
\begin{equation}
\chi_i=x^*_i\left(1+\frac{\lambda k_i \Theta^* \sum_{j=1}^n p_j k_j(1-x^*_j)^2}{\sum_{j=1}^n p_j k_j(x^*_j)^2}\right),
\end{equation}
which is Eq.~\eqref{eq:chi}.

\section{Conditions for $\mathbf{\mathcal{P}^o}$ to be nonempty} \label{appendix_D}

We test the validity of the three inequalities in \eqref{eq:po_cond}
in a sequential manner: First we study the condition for $\lambda$ when there 
exist $k_{i_1}$ and $k_{i_3}$ such that $g(k_{i_1},k_{i_3})\geq 0$ holds, we 
then test under the derived condition if there exists $k_{i_2}$ such that the 
other two inequalities hold.

As a preparation, we prove a result that will be used repeatedly in the rest 
of this appendix. In particular, we show that for $\mathcal{P}^o$ to be 
nonempty, it is necessary to have $\lambda z\leq 2$ from 
Eq.~\eqref{eq:po_prevalence}. Note that Eq.~\eqref{eq:po_prevalence} is the 
average of the function $f(k_a)=2k_a/(2+\lambda k_a)$ under the degree 
distribution $\mathbf{p}$. This function has a negative second order derivative 
$f^{\prime \prime}(k_a)=-8\lambda /(2+\lambda k_a)^3$, so $f(k_a)$ is concave. 
By Jensen's inequality, we have
\begin{equation}\label{eq:Jensen}
\sum_{i=1}^n p_i \frac{2k_i}{2+\lambda k_i}\leq \frac{2z}{2+\lambda z}.
\end{equation}
Since the left side equals $z/2$ from Eq.~\eqref{eq:po_prevalence}, it 
is necessary to have $2z/(2+\lambda z)\geq z/2$, which implies 
$\lambda z\leq 2$. The equality in Eq.~\eqref{eq:Jensen} holds only when $z$ 
is an integer and is one of the allowed degrees in $\mathbf{k}$ and, in 
addition, $\mathbf{p}$ concentrates on $z$. In this case we have $\lambda z=2$,
so $\mathcal{P}^o$ has a unique element $p$ that concentrates on $z$. 

We consider the case of $\lambda z<2$. The analysis begins with the setting
of the existence of $(k_{i_1},k_{i_3})$ such that $g(k_{i_1},k_{i_3})\geq 0$ 
holds. Defining $z^{+}=\mathrm{min}_i \{k_i\geq z\}$ and 
$z^{-}=\mathrm{max}_i \{k_i\leq z\}$, we have 
$k_{i_1}\in \{k_{1},k_2,\cdots,z^{+}\}$ and 
$k_{i_3}\in\{z^{-},\cdots,k_{n-1},k_n\}$. The function $g(k_a,k_b)$ is 
quadratic in $\lambda$, and the equation $g(k_a,k_b)=0$ has two roots: one 
positive and one negative. The positive one is 
\begin{equation}
\lambda^{(k_a,k_b)} =\frac{2}{z}-\frac{1}{k_a}-\frac{1}{k_b}+\sqrt{\left(\frac{1}{k_a}+\frac{1}{k_b}-\frac{2}{z}\right)^2+\frac{4}{k_a k_b}}. 
\end{equation}
As a result, for $0<\lambda<\lambda^{(k_a,k_b)}$, we have $g(k_a,k_b)<0$, 
whereas $g(k_a,k_b)\geq 0$ for $\lambda\geq \lambda^{(k_a,k_b)}$. If 
$\lambda^{(k_a,k_b)}$ is regarded as a function of $k_a$ and $k_b$, through
the derivatives, we have that $\lambda^{(k_a,k_b)}$ is a decreasing function 
of $k_a$ for $k_a>z$ and an increasing function of $k_b$ for $k_b<z$. 
Consequently, the value of $\lambda^{(k_a,k_b)}$ reaches its minimum at 
$(k_1,k_n)$. There exists at least one $(k_{i_1},k_{i_3})$ such that 
$g(k_{i_1},k_{i_3})\geq 0$ insofar as $\lambda\geq \lambda^{(k_1,k_n)}$.

Having determined the condition under which there exists $(k_{i_1},k_{i_3})$ 
such that $g(k_{i_1},k_{i_3})\geq 0$ holds, we can obtain the conditions 
under which there exists $k_{i_2}$ such that $g(k_{i_1},k_{i_2})\leq 0$ 
and $g(k_{i_2},k_{i_3})\leq 0$. When the curve $g(k_a,k_b)=0$ passes an 
integer point that can be chosen as $(k_{i_1},k_{i_3})$, we have 
$g(k_{i_1},k_{i_3})=0$. From Eq.~\eqref{eq:p3_formula}, we have $p_{i_2}=0$ and
\begin{equation}\label{eq:D3}
p_{i_1}=\frac{z-k_{i_3}}{k_{i_1}-k_{i_3}},\qquad p_{i_3}=\frac{k_{i_1}-z}{k_{i_1}-k_{i_3}}.
\end{equation}
We see that $p_{i_1}$ and $p_{i_3}$ are independent of the choice of $k_{i_2}$
and $\mathbf{p}$ is supported on one or two nodal degrees.

Now consider the case of $g(k_{i_1},k_{i_3})> 0$. For fixed $(k_{i_1},k_{i_3})$,
the inequalities $g(k_{i_1},k_{i_2})\leq 0$ and $g(k_{i_2},k_{i_3})\leq 0$ can 
be rearranged as
\begin{equation}\label{eq:ki2_ineq}
\begin{split}
&\left[\left(4\lambda-\lambda^2 z\right)k_{i_1}-2\lambda z\right]k_{i_2}\geq 2\lambda z k_{i_1}-4z,\\
&\left[\left(4\lambda-\lambda^2 z\right)k_{i_3}-2\lambda z\right]k_{i_2}\geq 2\lambda z k_{i_3}-4z.
\end{split}
\end{equation}
As $k_{i_1}\geq z$ and $\lambda z< 2$, we have
\begin{equation}
\left(4\lambda-\lambda^2 z\right)k_{i_1}-2\lambda z>0.
\end{equation}
From $g(k_{i_1},k_{i_3})\geq 0$ we have
\begin{equation}
\left(\left(4\lambda-\lambda^2 z\right)k_{i_3}-2\lambda z\right)k_{i_1}\leq 2\lambda z k_{i_3}-4z.
\end{equation}
Because $\lambda z < 2$ and $k_{i_3}\leq z$, the right-hand side of the above 
inequality is negative. We, thus, have
\begin{equation}
\left(4\lambda -\lambda^2 z\right)k_{i_3}-2\lambda z<0.
\end{equation}
With the above results, Eq.~\eqref{eq:ki2_ineq} implies that there exist 
feasible values of $k_{i_2}$ insofar as
\begin{equation}\label{eq:ki2_feasible}
\frac{2\lambda z k_{i_1}-4z}{ \left(4\lambda-\lambda^2 z\right)k_{i_1}-2\lambda z}\leq \frac{2\lambda z k_{i_3}-4z}{\left(4\lambda -\lambda^2 z\right)k_{i_3}-2\lambda z}
\end{equation}
and there is at least one integer between the two sides of the inequality.

Defining a function of $\lambda$ and $k_a$ as
\begin{equation}
f(\lambda, k_a)=\frac{2\lambda z k_a-4z}{\left(4\lambda-\lambda^2 z\right)k_a-2\lambda z},
\end{equation}
we have that the left and right sides of Eq.~\eqref{eq:ki2_feasible} are equal 
to $f(\lambda, k_{i_1})$ and $f(\lambda, k_{i_3})$, respectively. The 
derivative of $f(k_a)$ with respect to $k_a$ is
\begin{equation}\label{eq:df_ka}
\frac{\partial f(\lambda, k_a)}{\partial k_a}=
\frac{8 \lambda z \left(2-\lambda z\right)}{\left( \left(4\lambda-\lambda^2 z\right)k_a-2\lambda z\right)^2}.
\end{equation}
Consequently, $f(\lambda, k_a)$ is an increasing function of $k_a$ for 
$\lambda z<2$ and the function is non-singular, so $f(\lambda,k_{i_1})$ is 
bounded from above as
\begin{equation}\label{eq:lambda_sufficient_1}
f(\lambda, k_{i_1})<\lim_{k_a\to \infty} f(k_a)=\frac{2 z}{4-\lambda z}<z,
\end{equation}
whereas $f(\lambda, k_{i_3})$ is bounded from below as
\begin{equation}\label{eq:lambda_suffcient_2}
f(\lambda, k_{i_3})>\lim_{k_a\to 0} f(k_a)=\frac{2}{\lambda}>z.
\end{equation}
We thus have that the inequality $f(\lambda,k_{i_1})<f(\lambda, k_{i_3})$ 
holds for $\lambda z<2$. It remains to determine if there is an integer 
between $f(\lambda, k_{i_1})$ and $f(\lambda, k_{i_3})$. In this regard,
if $z$ is an integer and is one of the degrees allowed, the situation is 
relatively simple, and we pick $k_{i_2}=z$.

We analyze the case where $z$ is not an integer. Note that the left-hand 
side of Eq.~\eqref{eq:ki2_feasible} is strictly less than the right-hand side 
and $f(\lambda,k_a)$ is an increasing function of $k_a$. The gap between 
the two sides of Eq.~\eqref{eq:ki2_feasible} is then maximized for 
$(k_{i_1},k_{i_3})=(z^{+},z^{-})$. Suppose there are no integer points between 
$f(\lambda,z^{+})$ and $f(\lambda,z^{-})$. It implies that there are no 
integer points between $f(\lambda,k_{i_1})$ and $f(\lambda,k_{i_3})$ for any 
other choice of $(k_{i_1},k_{i_3})$. For $\lambda=\lambda^{(z^+,z^-)}$, the 
curve $g(k_a,k_b)=0$ passes the point $(k_a,k_b)=(z^{+},z^{-})$ and set 
$\mathcal{P}^o$ is nonempty based on Eq.~\eqref{eq:D3}, as we can take 
$(k_{i_1},k_{i_3})=(z^{+},z^{-})$. In the next, we show that if 
$\lambda>\lambda^{(z^+,z^-)}$, set $\mathcal{P}^o$ will be empty as there 
are no integer points between $f(\lambda,z^{+})$ and $f(\lambda,z^{-})$. 
However, for $\lambda<\lambda^{(z^+,z^-)}$, $\mathcal{P}^o$ is guaranteed to 
be nonempty.

To show that $\mathcal{P}^o$ is empty for $\lambda>\lambda^{(z^+,z^-)}$, we
note that the derivative of $f(\lambda,k_a)$ with respect to $\lambda$ is
\begin{equation}\label{eq:df_lambda}
\frac{\partial f(\lambda,k_a)}{\partial \lambda}=\frac{2z^2\left(\lambda k_a-2 \right)^2+16 z\left(k_a-z\right)}{\left[\left(4\lambda-\lambda^2 z\right)k_a-2\lambda z\right]^2}.
\end{equation}
For $k_a\geq z$ and $\lambda z<2$, the derivative is positive, so 
$f(\lambda,k_a)$ is an increasing function of $\lambda$. Now we show that if
\begin{equation}\label{eq:dn_cond}
 \left(4\lambda-\lambda^2 z\right)k_{a}-2\lambda z<0,
\end{equation}
then $f(\lambda,k_a)$ is a decreasing function of $\lambda$. Note that 
$k_{i_3}$ satisfies the above inequality for $g(k_{i_1},k_{i_3})>0$ [c.f., 
the discussions above Eq.~\eqref{eq:ki2_feasible}]. Taking the derivative 
with respect to $k_a$ for the numerator of the right-hand side of 
Eq.~\eqref{eq:df_lambda}, we get 
\begin{equation}
4\lambda z^2 (\lambda k_a-2)+16z>16z-8\lambda z^2>0,
\end{equation}
where the second inequality is the result of applying $\lambda z< 2$. The 
numerator on the right side of Eq.~\eqref{eq:df_lambda} itself is an 
increasing function of $k_a$. In addition, Eq.~\eqref{eq:dn_cond} implies 
\begin{equation}
k_a<\frac{2z}{4-\lambda z}.
\end{equation}
When $k_a$ equals the right side of this inequality, the numerator of the 
right side of Eq.~\eqref{eq:df_lambda} becomes,
\begin{equation}
\frac{16\lambda z^3 (\lambda z -2)}{(4-\lambda z)^2}<0,
\end{equation}
so $f(\lambda,k_a)$ is a decreasing function of $\lambda$ when 
Eq.~\eqref{eq:dn_cond} holds. For $\lambda=\lambda^{(z^+,z^-)}$, we have 
$f(\lambda,z^+)=z^{-}$ and $f(\lambda,z^-)=z^{+}$. For 
$\lambda>\lambda^{(z^+,z^-)}$, the left side of Eq.~\eqref{eq:ki2_feasible} 
increases from $z^-$ whereas the right side decreases from $z^+$. As a result, 
the gap between the two sides becomes smaller, and there cannot be any integer 
point in between.

We now show that, for $\lambda^{(k_1,k_n)}< \lambda<\lambda^{(z^+,z^-)}$, 
set $\mathcal{P}^o$ is guaranteed to be nonempty. In this region of 
$\lambda$, we have $g(k_1,k_n)>0 $ and $g(z^+,z^-)<0$. Consider the point 
$(k_a,k_b)=(z^+,k_n)$. For $g(z^+,k_n)=0$, according to Eq.~\eqref{eq:D3},
set $\mathcal{P}^o$ is nonempty. The other two possibilities: 
$g(z^+,k_n)>0$ and $g(z^+,k_n)<0$, can be treated separately. Suppose 
$g(z^+,k_n)>0$, we can pick $k_{i_1}=z^+, k_{i_2}=z^-$ and $k_{i_3}=k_n$. 
In this case, $g(k_{i_1},k_{i_2})<0$ and $g(k_{i_1},k_{i_3})>0$ hold by 
definition. It can then be shown that these two inequalities imply 
$g(k_{i_2},k_{i_3})<0$. In particular, note that
\begin{equation}\label{eq:D18}
\begin{split}
&g(k_{i_2},k_{i_3})-g(k_{i_1},k_{i_2})\\
=&\left(k_{i_1}-k_{i_3}\right)\left(\left(4\lambda-\lambda^2 z\right)k_{i_2}-2\lambda z\right).
\end{split}
\end{equation}
As $g({k_{i_1},k_{i_2}})< 0$, we have
\begin{equation}\label{eq:implicit_1}
\left(\left(4\lambda-\lambda^2 z\right)k_{i_2}-2\lambda z\right)k_{i_1}<2\lambda z k_{i_2}-4z.
\end{equation}
Since $k_{i_2}=z^{-}\leq z$ and $\lambda z<2$, the right side is negative and 
we have $\left(\left(4\lambda-\lambda^2 z\right)k_{i_2}-2\lambda z\right)<0$. 
This implies $g(k_{i_2},k_{i_3})<g(k_{i_1},k_{i_2})<0$. For the other case of 
$g(z^+,k_n)<0$, we pick $k_{i_1}=k_1$, $k_{i_2}=z^+$ and $k_{i_3}=k_n$, so
$g(k_{i_2},k_{i_3})<0$ and $g(k_{i_1},k_{i_3})>0$ hold by definition. From 
$\lambda z<2$ and $k_{i_2}=z^+\geq z$, we have 
$\left(\left(4\lambda-\lambda^2 z\right)k_{i_2}-2\lambda z\right)>0$. It can
thus be concluded from Eq.~\eqref{eq:D18} that 
$0>g(k_{i_2},k_{i_3})>g(k_{i_1},k_{i_2})$.

The above proof procedure can be applied to the case of picking 
$(k_{i_1},k_{i_2},k_{i_3})$ for 
$\lambda^{(k_1,k_n)}< \lambda<\lambda^{(z^+,z^-)}$. Suppose there are four 
degrees $k_a>k_b>z>k_c>k_d$, with $g(k_a,k_d)<0$ and $g(k_b,k_c)>0$. If the 
point $(k_b,k_d)$ has $g(k_b,k_d)>0$, we have that 
$(k_{i_1},k_{i_2},k_{i_3})=(k_b,k_c,k_d)$ defines a physical degree 
distribution from Eq.~\eqref{eq:p3_formula}. Similarly, if $g(k_a,k_c)<0$, 
we can choose $(k_{i_1},k_{i_2},k_{i_3})=(k_a,k_b,k_c)$.

The results of this appendix can be summarized as follows. Let 
$\lambda_2=\lambda^{(k_1,k_n)}$ and $\lambda_3=\lambda^{(z^+,z^-)}$. For 
$\lambda_1<\lambda<\lambda_2$, the set $\mathcal{P}^o$ is empty and we can 
only find the local maxima among all degree distributions that are supported 
on one or two nodal degrees. For $\lambda_2\leq \lambda \leq \lambda_3$, set $\mathcal{P}^o$ is nonempty, and it is necessary to further analyze if 
there are other local maxima supported on one or two nodal degrees and if 
the degree distributions in $\mathcal{P}^o$ are maxima and global maxima. 
For $\lambda>\lambda_3$, set $\mathcal{P}^o$ again becomes empty.

\section{Solution of the HMF equation for degree distribution supported on two nodal degrees} \label{appendix_E}

The equilibrium point $\mathbf{x}^*$ is given by the solution of
\begin{equation} \label{eq:equi_point}
\lambda k_{i_1}(1-x^*_{i_1})\Theta^*= x^*_{i_1},\ \lambda k_{i_2}(1-x^*_{i_2})\Theta^*= x^*_{i_2}.
\end{equation} 
Multiplying the two equations in Eq.~\eqref{eq:equi_point} by $p_{i_1}$ and 
$p_{i_2}$, respectively, and summing them, we obtain
\begin{equation}\label{eq:D_local0}
\psi(\mathbf{p})=\lambda z\Theta^* -\lambda z\left(\Theta^*\right)^2. 
\end{equation}
To obtain $\psi(\mathbf{p})$, it suffices to find the value of $\Theta^*$.

Equation~\eqref{eq:equi_point} gives
\begin{equation}\label{eq:D_local}
x^*_{i_1}=\frac{\lambda k_{i_1} \Theta^*}{1+\lambda k_{i_1}\Theta^*},\ x^*_{i_2}=\frac{\lambda k_{i_2} \Theta^*}{1+\lambda k_{i_2}\Theta^*}.
\end{equation}
From the definition of $\Theta^*$, we have 
\begin{equation} \label{eq:D_zTheta}
\begin{split}
z\Theta^*=p_{i_1} k_{i_1} x^*_{i_1}+p_{i_2} k_{i_2} x^*_{i_1}.
\end{split}
\end{equation}
Substituting Eq.~\eqref{eq:D_local} and the values $p_{i_1}$ and $p_{i_2}$ in 
Eq.~\eqref{eq:p_ab} into Eq.~\eqref{eq:D_zTheta}, we obtain the following 
quadratic equation for $\Theta^*$:
\begin{equation} \label{eq:D_local_2}
\beta_2 \left(\Theta^*\right)^2+\beta_1 \Theta^* +\beta_0=0,
\end{equation}
with the coefficients, 
\begin{equation}
\begin{split}
&\beta_2=\lambda^2 z k_{i_1} k_{i_2}, \\
&\beta_1=\lambda z \left( k_{i_1}  +k_{i_2}  - \lambda k_{i_1} k_{i_2} \right),\\
&\beta_0=z - \lambda z k_{i_1} - \lambda z k_{i_2}  +  \lambda k_{i_1} k_{i_2}.
\end{split}
\end{equation}
Noting that the second moment of the degree distribution is
\begin{equation} \label{eq:second_moment}
\langle k^2\rangle=p_{i_1} k_{i_1}^2+p_{i_2} k_{i_2}^2=z\left(k_{i_1}+k_{i_2}\right)-k_{i_1}k_{i_2},
\end{equation}
We have $\beta_0=z-\lambda\langle k^2\rangle<0$ as $\lambda$ is above the 
epidemic outbreak threshold $z/\langle k^2\rangle$. Since $\beta_2>0$, 
the only physical solution (with $0<\Theta^*<1$) of Eq.~\eqref{eq:D_local_2} is
\begin{equation} \label{eq:D_Theta}
\Theta^*=\frac{-\beta_1+\sqrt{\beta_1^2-4\beta_2\beta_0}}{2\beta_2}.
\end{equation}
Substituting Eq.~\eqref{eq:D_Theta} into Eq.~\eqref{eq:D_local0}, we obtain 
the value of $\psi(\mathbf{p})$ as given by Eq.~\eqref{eq:psi_two}.


%
\end{document}